\documentclass [11pt,a4paper]{article}
\pdfoutput=1 
\usepackage{jheppub}
\usepackage{latexsym,epsfig,amssymb, amsmath,nicefrac}
\usepackage{xcolor}
\usepackage{subcaption}

\usepackage{slashed}

\usepackage{tikz}
\usepackage{comment}
\usepackage{color}
\usepackage[makeroom]{cancel}
  \usepackage{latexsym}
  \usepackage{epsf}
  \usepackage{amssymb}
  \usepackage{graphicx}
  \usepackage{amsmath}
  \usepackage{amsmath,amssymb,amsthm}
  \usepackage{verbatim}
  \usepackage{hyperref}
\usepackage{physics}
\usepackage{float}

\def\tb0{\tilde{\beta}_0}
\def\b0{\beta_0}

\def\bi{\begin{itemize}}
\def\ei{\end{itemize}}
\def\be{\begin{equation}}
\def\ee{\end{equation}}
\newcommand{\bea}{\begin{eqnarray}}
\newcommand{\eea}{\end{eqnarray}}
 \renewcommand{\i}{{\mathrm{i}}}

\renewcommand{\Re}{\textrm{Re}\,}

\def\Kahler{K\"{a}hler~}

\newcommand{\vol}{\mathcal{V}}
\renewcommand{\v}{\vol}
\newcommand{\p}{\partial}
\renewcommand{\b}[1]{\left(#1\right)}
\renewcommand{\sb}[1]{\left[#1\right]}
\renewcommand{\L}{\mathcal{L}}
\newcommand{\X}{\mathcal{X}}
\newcommand{\N}{\mathcal{N}}

\begin{document}

\begin{flushright}
MPP-2023-31
\end{flushright}

\title{Early Dark Energy in Type IIB String Theory}
\author[a,b]{Michele Cicoli,}
\author[a,b]{Matteo Licheri,}
\author[b]{Ratul Mahanta,}
\author[c]{Evan McDonough,}
\author[a,b]{Francisco G. Pedro,}
\author[d]{Marco Scalisi}

\affiliation[a]{\footnotesize Dipartimento di Fisica e Astronomia, Universit\`a di Bologna, via Irnerio 46, 40126 Bologna, Italy}
\affiliation[b]{\footnotesize INFN, Sezione di Bologna, viale Berti Pichat 6/2, 40127 Bologna, Italy}
\affiliation[c]{\footnotesize 
Department of Physics, University of Winnipeg, Winnipeg MB, R3B 2E9, Canada}
\affiliation[d]{\footnotesize Max-Planck-Institut f\"ur Physik (Werner-Heisenberg-Institut), F\"ohringer Ring 6, 80805, M\"unchen,
Germany}

\emailAdd{michele.cicoli@unibo.it}
\emailAdd{matteo.licheri@unibo.it}
\emailAdd{mahanta@bo.infn.it}
\emailAdd{e.mcdonough@uwinnipeg.ca}
\emailAdd{francisco.soares@unibo.it}
\emailAdd{mscalisi@mpp.mpg.de}

\abstract{Early Dark Energy (EDE) is a promising model to resolve the Hubble Tension, that, informed by Cosmic Microwave Background data, features a generalization of the potential energy usually associated with axion-like particles. We develop realizations of EDE in type IIB string theory with the EDE field identified as either a $C_4$ or $C_2$ axion and with full closed string moduli stabilization within the framework of either KKLT or the Large Volume Scenario. We explain how to achieve a natural hierarchy between the EDE energy scale and that of the other fields within a controlled effective field theory. We argue that the data-driven EDE energy scale and decay constant can be achieved without any tuning of the microscopic parameters for EDE fields that violate the weak gravity conjecture, while for states that respect the conjecture it is necessary to introduce a fine-tuning. This singles out as the most promising EDE candidates, amongst several working models, the $C_2$ axions in LVS with 3 non-perturbative corrections to the superpotential generated by gaugino condensation on D7-branes with non-zero world-volume fluxes.}

\maketitle

\section{Introduction}
\label{Intro}

The Hubble constant $H_0$, as inferred from Planck 2018 Cosmic Microwave Background (CMB) data \cite{Aghanim:2018eyx}, is in $5\sigma$ disagreement with the SH0ES  cosmic distance ladder measurement \cite{Riess:2021jrx}. This `Hubble tension'  has spurred on an intense experimental effort and the development of new ways to measure $H_0$ (see \cite{Kamionkowski:2022pkx,Verde:2019ivm} for reviews). The tension persists between varied early and late universe probes at the level of $4$-$6\sigma$ \cite{Verde:2019ivm}.
A commensurate effort has been made on the theory side, aimed at developing an alternative cosmological model to bring these measurements in agreement. Amongst the theory approaches, the modification of early universe physics holds particular promise (see \cite{Knox:2019rjx}) by satisfying first and foremost the tight constraints that the CMB places on any new cosmological physics. A detailed review is provided in Sec.~\ref{sec:EDEreview} (see also the review section of \cite{McDonough:2022pku}).

Early Dark Energy (EDE) \cite{Poulin:2018cxd,Poulin:2023lkg} is an example of new physics in the early universe that resolves the Hubble tension by bringing the CMB inference into agreement with SH0ES, while leaving the former nearly indistinguishable from $\Lambda$CDM. The model proposed in \cite{Poulin:2018cxd}  utilizes a scalar field with potential energy $V(\varphi) = V_0 \left[1- \cos \left(\varphi/f\right)\right]^3$ 
, featuring an exponent that distinguishes it from the conventional potential of an axion-like particle. This potential is {\it motivated by data}: it provides a significantly better fit to the data than a monomial $V\sim \varphi^{2n}$ \cite{Smith:2019ihp} 
or a cosine with a different exponent \cite{Poulin:2018cxd}. The vast majority of work on EDE (see e.g.~\cite{Ivanov:2020ril,Hill:2020osr,McDonough:2022pku}) has therefore focused on this form of the potential, though alternative EDE-like models abound \cite{Kaloper:2019lpl,Agrawal:2019lmo,Alexander:2019rsc,Lin:2019qug,Niedermann:2019olb,Berghaus:2019cls,Sakstein:2019fmf,Ye:2020btb,Niedermann:2020dwg,Seto:2021xua,Alexander:2022own,Rezazadeh:2022lsf,McDonough:2021pdg,Lin:2022phm,Burgess:2021obw}. This work has elucidated challenges to the model from data, in particular, tension with large scale structure (see e.g.~\cite{Ivanov:2020ril,Hill:2020osr}), that has motivated extensions of the EDE model, see ~\cite{McDonough:2021pdg,Clark:2021hlo,Allali:2021azp,Ye:2021iwa}, to include an additional ultralight axion dark matter component \cite{Allali:2021azp,Ye:2021iwa}. Relatively little input has come from the formal theory community, with exception of Refs.~\cite{McDonough:2022pku} and \cite{Rudelius:2022gyu}. 

In this work, we seek to identify and address the challenges to building a phenomenologically viable EDE model within the context of string theory. The first steps have been already provided in \cite{McDonough:2022pku}, in the context of KKLT compactifications, with the EDE field identified as a $C_2$ axion. Its potential is derived from non-perturbative corrections to the superpotential $W$ generated by gaugino condensation on D5-branes. Besides the need to tune the prefactors of these non-perturbative effects to reproduce the correct EDE scale, it remains unclear if gaugino condensation on D5-branes can actually yield a non-zero contribution to the superpotential for cycles in the geometric regime\footnote{See however \cite{Ben-Dayan:2014lca} for cosmological applications.} \cite{McAllister:2008hb, Cicoli:2021tzt}.
 
Here we go beyond what achieved in \cite{McDonough:2022pku} and perform a deeper analysis of EDE model building in type IIB string theory which is one of the most promising corners of string theory for moduli stabilization. We propose string embeddings of EDE in the moduli stabilization frameworks of KKLT \cite{Kachru:2003aw} and the Large Volume Scenario (LVS) \cite{Balasubramanian:2005zx, Cicoli:2008va}. Moreover, we identify different choices of axion as the EDE candidate. In particular, we try to realize the EDE potential $V= V_0 \left[1- \cos(\varphi/f)\right]^3$ with the phenomenologically relevant parameters $V_0 \sim {\rm eV}^4$ and $f\simeq 0.2\, M_P$, while satisfying the following conditions:

\begin{enumerate}
\item {\bf Controlled de Sitter moduli stabilization:} All string moduli should be stabilized in a dS vacuum where the effective field theory is under control. In particular the compactification volume should be large enough to trust the $\alpha'$ expansion, the string coupling should be small enough to remain in the perturbative regime, and the instanton expansion should be well behaved. 
One of the main obstacles against achieving moduli stabilization with full control is the fact that the decay constant $f$ of the EDE field has to be relatively close to the Planck scale. This can intuitively be seen as follows. Explicit string computations \cite{Svrcek:2006yi, Conlon:2006tq, Cicoli:2012sz}, as well as the weak gravity conjecture applied to axions \cite{Arkani-Hamed:2006emk, Rudelius:2015xta, Brown:2015iha, Hebecker:2015zss}, give $f S \simeq \lambda M_P$ where $S$ is the instanton action and $\lambda$ an $\mathcal{O}(1)$ constant. Hence, $f\simeq 0.2\, M_P$ implies $S\sim \mathcal{O}(10)$. Given that in string compactifications $S$ is set by the volume of the internal cycle wrapped by the instanton which generates the EDE potential, $f\simeq 0.2\, M_P$ implies volumes of $\mathcal{O}(10)$ in string units which might not be large enough to control the effective field theory.

\item \textbf{Decoupling of non-EDE modes:} All moduli different from the EDE field should be stabilized at an energy scale much larger than $V_0$, so that they become much heavier than the EDE field whose mass is of order $m\sim 10^{-27}$ eV. This requirement is needed for two reasons: ($i$) saxions with masses below about $1$ meV would mediate unobserved fifth-forces; ($ii$) the dynamics of ultra-light axions with masses around $m\sim 10^{-27}$ eV could play a significant role around matter-radiation equality, potentially modifying the cosmological evolution of the EDE model\footnote{On the other hand, an ultra-light axion component of dark matter may in fact help the model to be in agreement with Large Scale Structure data \cite{Allali:2021azp,Ye:2021iwa}.}. The EDE scale $V_0$ should also be decoupled from the scale of supersymmetry breaking and the gravitino mass.

\item {\bf Absence of fine-tuning:} The main phenomenological features of the model, namely the desired EDE energy scale $V_0$, the typical $[1- \cos(\varphi/f)]^3$ shape of the potential and the decoupling of the non-EDE modes, should be realized without the need to fine-tune the underlying microscopic parameters. If instead some parameters need to take unnatural values, the UV completion should provide enough tuning freedom.
  
\item {\bf Explicit Calabi-Yau realization:} A full-fledged string model of EDE should feature a globally consistent compactification with an explicit Calabi-Yau orientifold involution and brane setup which allow for tadpole cancellation and the realization of all perturbative and non-perturbative ingredients needed to fix the moduli and generate the EDE potential.
\end{enumerate}
These challenges are not independent, but instead exhibit a rich interplay. For example, large volume can help achieve a convergent $\alpha'$ and instanton expansion and the desired EDE energy scale, but at the cost of lowering both the decay constant and all mass scales. On the other hand, at moderate volume, the EDE energy scale can be adjusted simply by lowering the prefactors of non-perturbative terms (this was the approach of \cite{McDonough:2022pku}), at the cost of introducing an exponential fine-tuning in the model.

We elucidate and address the first 3 challenges within the context of various string theory realizations of EDE. We consider moduli stabilization both in the context of KKLT and LVS (a concise summary of the string theory background is provided in Sec.~\ref{sec:2stringreview}). These scenarios have been studied in depth in the literature and are two of the most promising frameworks for controlled moduli stabilization. 

In order to ensure the decoupling of non-EDE modes, we consider both $C_4$ and $C_2$ axions as potential EDE candidates. In fact, both of these fields enjoy a continuous shift symmetry, which is exact at the perturbative level. Hence, any perturbative correction would generate the required hierarchy by fixing the corresponding saxions while leaving the $C_4$ and $C_2$ axions flat. This is what happens typically for $C_2$ axions and the bulk $C_4$ axion in LVS. The situation is somewhat different for $C_4$ axions in KKLT where moduli stabilization relies only on non-perturbative effects. In this case the $C_4$ axion cannot therefore play the role of the EDE field. 

Regarding the task to reproduce the EDE scale, $C_2$ axions seem more promising than $C_4$ axions. An intuitive explanation concerns the fact that matching $V_0 \sim {\rm eV}^4$ without fine-tuning requires a violation of the weak gravity conjecture applied to axions. In fact, writing again $f S \simeq \lambda M_P$, the EDE scale can be written as
\begin{equation}
V_0 = A\,e^{-S}\,M_P^4 \simeq A\,e^{- \lambda M_P/f}\,M_P^4 \simeq A\,e^{- 5 \lambda}\,M_P^4\qquad \text{for}\qquad f \simeq 0.2\,M_P\,.
\label{KeyRel}
\end{equation}
Demanding $V_0 \sim {\rm eV}^4\sim 10^{-108}\,M_P^4$ corresponds to $A\,e^{- 5 \lambda}\sim 10^{-108}$, which clearly requires $A\ll 1$ for $\lambda\sim\mathcal{O}(1)$.\footnote{Here our logic is different from the one of \cite{Rudelius:2022gyu}, which set $A\simeq 10^{-8}$ to get an overall $M_{\rm GUT}^4$ scale and fixed $\lambda\sim\mathcal{O}(1)$ to infer the value of $f$ needed to match the EDE scale. This logic yields $f \simeq 0.008\,M_P$, which is below the best fit value $f \simeq 0.2\,M_P$.} On the other hand, cases with $\lambda \gg 1$ (i.e. violation of the weak gravity conjecture) could reproduce the correct EDE scale without the need to tune the prefactor $A$. Ref. \cite{Cicoli:2021gss} found $\lambda\sim\mathcal{V}^{1/3}\gg 1$ (where $\mathcal{V}\gg 1$ is the compactification volume in string units) for $C_2$ axions with superpotential from fluxed ED3-instantons/gaugino condensation on D7s, while $\lambda\sim\mathcal{O}(1)$ for $C_2$ axions with ED1-instanton corrections to the K\"ahler potential, and $C_4$ axions with superpotential generated by ED3-instantons/gaugino condensation on D7-branes. Therefore, we identify $C_2$ axions, with a potential generated by fluxed ED3-instantons or gaugino condensation on D7-branes with non-zero world-volume fluxes \cite{Cicoli:2021gss}, as in principle the most promising candidates to match the EDE scale with minimal fine-tuning of the model.

However, as we shall see, matching the required EDE scale without any tuning of the prefactors of non-perturbative effects leads to a compactification volume of order $\v\sim\mathcal{O}(10^4$-$10^5)$. In turn, obtaining $f\simeq 0.2\,M_P$ requires $\mathcal{O}(100)$ D7-branes supporting the gaugino condensates which generate the EDE potential. This number is relatively large but still achievable in F-theory compactifications \cite{Louis:2012nb}. Moreover, $\v\sim\mathcal{O}(10^4$-$10^5)$ can be obtained easily in LVS models, while in KKLT it requires scenarios with $\mathcal{O}(1000)$ D7-branes supporting the gaugino condensate that yields the leading KKLT potential, otherwise the mass of the volume modulus would be below the cosmological moduli problem bound, $m_\v \lesssim 50$ TeV. Such a large number of D7-branes might be hard to achieve in a way compatible with D7 tadpole cancellation and a controlled backreaction. Thus, our analysis indicates that the most promising candidates to realize EDE from type IIB string theory are $C_2$ axions in LVS with a potential generated by gaugino condensation on D7-branes with non-vanishing gauge fluxes. Fluxed ED3-instantons would instead not be compatible with $f\simeq 0.2\,M_P$ for $\v\sim\mathcal{O}(10^4$-$10^5)$. 

The main challenge left is to construct an explicit Calabi-Yau embedding of these models where all the needed non-perturbative effects are explicitly shown to arise with the exact coefficients needed to reproduce the $[1- \cos(\varphi/f)]^3$ shape of the EDE potential.

Let us also point out that the KKLT and LVS implementations of EDE are distinguished in part by the mass of the gravitino. The most natural LVS models predict a gravitino mass far above the energy scale of any particle physics experiment. In particular, typical Swiss-cheese Calabi-Yau models lead to $m_{3/2}\sim \mathcal{O}(10^{13})$ GeV, while K3-fibered compactifications feature $m_{3/2}\sim \mathcal{O}(10^{10})$ GeV. On the other hand, KKLT models with the lowest possible number of D7-branes correlate with a TeV-scale gravitino mass. This suggests that experimental searches for the gravitino may be complementary to cosmological searches for EDE as it emerges from KKLT. An additional complementary direction is to interface the LVS EDE models with LVS inflation models \cite{Conlon:2005jm,Cicoli:2008gp,Cicoli:2011ct,Broy:2015zba,Cicoli:2016chb, Cicoli:2016xae, Cicoli:2017axo}, wherein the volume ${\cal V}$ is fixed by matching to the amplitude of the CMB power spectrum, even if in some models the volume can evolve from inflation to today \cite{Conlon:2008cj, Cicoli:2015wja}.

These analyses suggest that EDE can be a viable cosmological model from the perspective of string theory. The more difficult model building task is to realize multiple non-perturbative contributions to $W$ with precise coefficients that reproduce the EDE potential, even if other features of the model (like moduli stabilization, the EDE scale and decay constant, and the decoupling of non-EDE modes) can be achieved in LVS without fine-tuning. Hence we do not consider the theory challenges so different in difficulty in comparison to those faced by other cosmological models, such as dark energy \cite{Choi:1999xn,Cicoli:2012tz,Cicoli:2018kdo,Olguin-Trejo:2018zun,Cicoli:2021fsd,Cicoli:2021skd}, or fuzzy dark matter \cite{Hui:2016ltb, Cicoli:2021gss}, and thus one expects an eventual plethora of model realizations, of which we have only scratched the surface. By expanding the playground of model-building frameworks for EDE, this work will enable future efforts to target specific aspects of phenomenology that may be of observational interest, such as the coupling to photons \cite{Cicoli:2012sz} and the associated particle and gravitational wave production \cite{Weiner:2020sxn}.

Notation: In this work $M_P = 2.435 \times 10^{18}$ GeV is the reduced Planck mass.

\section{Early Dark Energy and the Hubble Tension}
\label{sec:EDEreview}

The Hubble tension is sharpest between Planck 2018 CMB data and SH0ES cosmic distance ladder measurement. Here we focus on these two experiments, but re-emphasize that the tension exists between varied data sets -- see \cite{Kamionkowski:2022pkx} for a recent review. What follows is intended to be a non-technical review of the essential physics of the Hubble tension and the EDE model (see \cite{Poulin:2023lkg} for a detailed review), with a particular focus on the constraints from data that guide the model-building process. This is complementary but distinct from the review given in \cite{McDonough:2022pku}. 

Key to understanding the EDE approach to the Hubble tension is that the CMB data, namely the distribution of CMB anisotropies on the sky, is an intrinsically two-dimensional picture of the universe. Thus, while one may directly measure the {\it angular} scale of features in the CMB, to translate this into length scales one must assume a cosmological model. The Hubble length, $H_0 ^{-1}$, is one such length scale that one may try to infer.

Indeed the most precise cosmological measurement to date is the Planck 2018 measurement of the angular extent of the comoving sound horizon at last scattering, $100\ \theta_s = 1.0411 \pm 0.0003$ \cite{Planck:2018vyg}.   This is defined by a ratio of length scales, as
\begin{equation}
\label{eq:thetas}
    \theta_s = \frac{r_s (z_*)}{D_A(z_*)}\,,
\end{equation}
where $r_s$ measures distances between points {\it in} the surface of last scattering\footnote{Last scattering surface refers to the time of last scattering of photons and electrons before the recombination of electrons and protons into hydrogen. For a review of CMB physics and terminology, see \cite{Hu:2001bc}.}, while $D_A$ corresponds to the distance from an observer {\it to} the CMB last scattering surface. More precisely, $r_s(z_*)$ is the comoving sound horizon at last scattering, defined as
\begin{equation}
\label{eq:rs}
   r_s(z_*) = \int _{z_*} ^{z_{\rm re}} \frac{{\rm d} z}{H(z)} c_s(z) ,
\end{equation}
with $z_*$ the redshift of last scattering, $z_{\rm re}$ is the redshift of reheating after cosmic inflation, and $c_s$ the sound speed of the photon-baryon plasma, whereas $D_A(z_*)$ is the angular diameter distance to the surface of last scattering,
\begin{equation}
D_A (z_*) = \int _0 ^{z_*} {\rm d}z \frac{1}{H(z)}\,,
\end{equation}
which is sensitive to $H(z=0)$, i.e. the Hubble constant $H_0$. These expressions suggest a path forward for resolving the Hubble tension: The $0.03\%$ measurement of the angle $\theta_s$ can accommodate the $\approx 10\%$ increase in $H_0$  if there is a commensurate increase in $H(z\sim z_*)$. This approach, which acts to reduce the sound horizon at last scattering, has been extensively studied (see \cite{Kamionkowski:2022pkx} for a review). A popular model realization is Early Dark Energy \cite{Poulin:2018cxd}.

The reduction of the sound horizon can be easily achieved by an ultralight scalar, satisfying the Klein-Gordon equation,
\begin{equation}
    \ddot{\varphi}+3 H \dot{\varphi} + V'=0\,.
\end{equation}
At early times, when the Hubble drag term dominates the dynamics, the scalar is nearly frozen in place and contributes a dark energy-like component to the universe. This phase eventually terminates, as the contents of the universe redshift and the Hubble parameter decreases, releasing the field from Hubble drag and triggering the decay of the EDE. This occurs around the time at which $H^2 \sim V'' $. The decay of the EDE is necessary to avoid any unintended impact on post-CMB physics. On the other hand, in order to have any sizeable effect on the sound horizon, the decay of the EDE must happen within the decade of redshift preceding last scattering \cite{Knox:2019rjx}. This fixes the mass of the ultralight scalar to $m \sim 10^{-27}\, {\rm eV}$.

The sound horizon is not the only scale probed the CMB, and likewise the dark energy -like phase of the EDE is not the only aspect of the dynamics that is constrained by data. The dissipation of CMB anisotropies on small angular scales (high multipole moment $\ell$), known as `Silk Damping',  provides another characteristic scale -- the damping scale $r_d$. The damping scale constrains the decay of the EDE via its impact on the relative size $r_s/r_d$. CMB data selects as the best EDE-like model the one that maximizes the decrease in $r_s$ and minimizes the change in $r_s/r_d$ \cite{Poulin:2018cxd}.

Putting these puzzle pieces together, one may build a concrete model. A well studied example is given by \cite{Poulin:2018cxd}
\be
V(\varphi) = V_0 \left[ 1- \cos\left(\frac{\varphi}{f}\right)\right]^3
= V_0\left[\frac52-\frac{15}{4}\cos\left(\frac{\varphi}{f}\right)+\frac32\cos\left(\frac{2\varphi}{f}\right)-\frac14\cos\left(\frac{3\varphi}{f}\right)\right],
\label{eq:EDE_V}
\ee
with $V_0 \equiv m^2 f^2$. The EDE potential (\ref{eq:EDE_V}) may be thought of as a generalization of the usual axion potential. The unconventional exponent is selected by data, which can be understood as largely due to the ability of the model to reduce the sound horizon while minimizing the impact on the damping scale, as described above. The exponent determines the shape of the potential near the minimum as locally $V \propto \varphi^6$, such that the energy density redshifts as
$a^{-9/2}$ in the decaying phase. This can be contrasted with the conventional axion potential, $V \sim 1-\cos(\varphi/f)$, which has a quadratic minimum, leading to a dark matter-like evolution in the decaying phase. An ultra-light axion component of dark matter is tightly constrained by data \cite{Hlozek:2014lca,Lague:2021frh} and can not resolve the Hubble tension. This model can also be contrasted with a monomial EDE potential $V= V_0 \left(\varphi/M_P\right)^{2n}$ \cite{Agrawal:2019lmo}, which, due to the convexity of the potential and the dynamics of perturbations, is strongly disfavored by CMB data relative to a cosine-type potential \cite{Smith:2019ihp}. We note that similar generalizations of an axion potential have been studied as an inflation model in ~\cite{Higaki:2015kta,Czerny:2014wza,Czerny:2014xja}.

The parameter values relevant to the Hubble tension in the EDE model, eq.~\eqref{eq:EDE_V}, follow from simple considerations. Electrons and protons recombine when the temperature of the primordial plasma drops below $T \sim $ eV, selecting $V_0\sim {\rm eV}^4$ as the benchmark energy scale if the EDE is to play a cosmologically relevant role around that time. The mass of the EDE scalar field should be comparable to the Hubble parameter at that time, $H \sim T^2/M_P$, 
to trigger the decay of the EDE, which fixes $m\sim 10^{-27}$ eV. From \eqref{eq:EDE_V}, these determine the decay constant as $f\sim M_P$. These order of magnitude estimates are born out in the fit to data, which selects out a near- but sub-Planckian decay constant, $f\simeq 0.2\, M_P$, as the preferred value \cite{McDonough:2021pdg}. 

The EDE model is a promising candidate to replace $\Lambda$CDM as the concordance  model of cosmology \cite{DODELSON20211}. However, the model faces serious challenges from both theory and data, as discussed in the introduction, that bring this privileged status into question \cite{Hill:2020osr}. On the data side, chief among these is the tension of EDE with large scale structure (LSS) data \cite{Hill:2020osr,Ivanov:2020ril,DAmico:2020ods,Jedamzik:2020zmd,Lin:2021sfs,Goldstein:2023gnw} (see also \cite{Smith:2020rxx,Murgia:2020ryi,Simon:2022adh,Herold:2021ksg}). As discussed in detail in \cite{Hill:2020osr,Ivanov:2020ril,DAmico:2020ods,Jedamzik:2020zmd,Lin:2021sfs} and reviewed in \cite{McDonough:2022pku},  the addition of an EDE-like component necessitates a commensurate increase in the amount of dark matter, to compensate the impact of the EDE on the redshifting of CMB photons, as encoded in the height of the first peak of the CMB temperature anisotropy angular power spectrum. This increased dark matter is in tension with observations of weak gravitational lensing and galaxy clustering \cite{Hill:2020osr,Ivanov:2020ril}, such as data from the Dark Energy Survey \cite{DES:2021wwk}, and from BOSS \cite{Alam:2016hwk}. The tension with LSS can be ameliorated by adding in additional degrees of freedom, such as in \cite{McDonough:2021pdg, Clark:2021hlo,Allali:2021azp,Ye:2021iwa}. We also note the preference for a non-zero EDE component from the Atacama Cosmology Telescope, see Ref.~\cite{Hill:2021yec,Poulin:2021bjr}. These results add to the motivation to study the EDE in a UV complete framework, such as string theory.

\section{Moduli Stabilization}
\label{sec:2stringreview}

We now shift gears to string theory. We first review the low energy effective field theory of the KKLT \cite{Kachru:2003aw} and LVS \cite{Balasubramanian:2005zx,Cicoli:2008va} approaches to moduli stabilization. These frameworks provide in fact ideal settings for a controlled EDE dynamics in string theory.

We will hereby assume that the axion-dilaton and complex structure moduli are stabilized at a higher scale, and that all quantities, implicitly depending on these fields, such as the flux-induced superpotential $W_0$ or the prefactor of instanton corrections,  can be regarded as constant. The F-term scalar potential is then calculated employing the supergravity formula (setting $M_P=1$)
\begin{equation}
\label{eqn:SUGRApotential}
    V = e^{K}\b{D_{I} W K^{I\bar{J}} D_{\bar{J}} \overline{W}-3|W|^2} \, ,
\end{equation}
where $I$ labels chiral superfields, the covariant derivative is $D_IW \equiv \partial_I W + W \partial_I K$ and $K^{I\bar{J}}$ is the inverse of the \Kahler metric $K_{I\bar{J}}\equiv \partial_I \partial_{\bar{J}}K$. Moreover, the gravitino mass is defined as
\begin{equation}
    m_{3/2} = e^{K/2}|W| \, .
\end{equation}
Let us note that, in the present paper, we will focus on type IIB compactifications, where the tree-level K\"ahler potential reads
\begin{equation}
    K= K_{\rm K\ddot{a}hler}+K_{\rm cs}+K_{\rm dilaton}\,,
\end{equation}
and one has that the overall factor, both in the gravitino mass and in the F-term potential, factorizes as $e^K=e^{K_{\rm K\ddot{a}hler}}e^{K_{\rm cs}}e^{K_{\rm dilaton}}$. Given that we will focus exclusively on the K\"ahler moduli sector, we will henceforth omit the $e^{K_{\rm cs}}e^{K_{\rm dilaton}}$ factors, though the reader should be aware that they are implicit throughout. In our estimates of the energy scales relevant for phenomenology, we will simply set $e^{K_{\rm cs}}e^{K_{\rm dilaton}}=1$.

\subsection{KKLT}

Let us first focus on KKLT \cite{Kachru:2003aw}. We consider a simple model consisting in only one \Kahler modulus $T = \tau + i \theta$ and a chiral nilpotent superfield $X$ \cite{Casalbuoni:1988xh,Komargodski:2009rz,Volkov:1973ix,Ivanov:1978mx,Rocek:1978nb,Lindstrom:1979kq,Antoniadis:2014oya,Ferrara:2014kva,Kallosh:2014hxa,Scalisi:2015qga}, which encapsulates the degrees of freedom of an anti-D3 brane providing the uplift mechanism \cite{Kallosh:2014wsa, Bergshoeff:2015jxa, Kallosh:2015nia,Polchinski:2015bea,Garcia-Etxebarria:2015lif,Bandos:2015xnf,Dasgupta:2016prs,Vercnocke:2016fbt,Kallosh:2016aep,Bandos:2016xyu,GarciadelMoral:2017vnz,Kallosh:2018wme,Kallosh:2018psh,Kallosh:2019axr,Cribiori:2019hod,Parameswaran:2020ukp}. The nilpotency condition $X^2=0$ implies that the scalar part of the superfield $X$ is effectively replaced by a fermion bi-linear. Since the expectation value of this is expected to be zero (except if the fermion condenses in the vacuum), we can safely set $X=0$ in the final output of the scalar potential. Let us stress that in this paper we focus on anti-D3 brane uplifting for the sake of simplicity but all our considerations on EDE from string theory are completely independent from the mechanism responsible to achieve a dS minimum, and so would apply more in general (for other uplifting scenarios see for example \cite{Cicoli:2015ylx} for T-branes, \cite{Westphal:2006tn} for $\alpha'$ effects, \cite{Gallego:2017dvd} for non-zero F-terms of the complex structure moduli and \cite{Cicoli:2012fh} for non-perturbative effects at singularities). 

The \Kahler potential in KKLT may be written as (with $\v= (2\tau)^{3/2}$)
\begin{equation}
K = -3 \ln \b{T+\bar{T}} +3\,\frac{X \bar{X}}{T+\bar{T}}\,,
\end{equation}
while the superpotential reads
\begin{equation}\label{eqn:W_KKLT}
    W = W_0 + M X + A\, e^{-\mathfrak{a} T} \, .
\end{equation}
Writing $W_0=|W_0|\,e^{i\phi_{W_0}}$ and $A=|A|\,e^{i\phi_A}$, without loss of generality, we set $\phi_{W_0}=\pi$ and $\phi_A=0$, so that $W_0 = -|W_0|$ and $A = |A| \in \mathbb{R}^+$.
Using \eqref{eqn:SUGRApotential} one can compute the uplifted scalar potential 
\begin{equation}
  V_\text{KKLT} = \frac{\mathfrak{a}^2 A^2 e^{-2 \mathfrak{a} \tau}}{6 \tau}+\frac{\mathfrak{a} A^2 e^{-2 \mathfrak{a} \tau}}{2 \tau^2}-\frac{\mathfrak{a}  A |W_0| e^{-\mathfrak{a} \tau}}{2 \tau^2}\cos(\mathfrak{a} \theta) +\frac{M^2}{12 \tau^2} \, ,
\label{VKKLT}
\end{equation}
where the nilpotency condition has been imposed. The minimum for the axion lies at the origin: $\theta=0$. Note that a different choice of the phase of $W_0$ would give a different location of the axion minimum. For example, choosing $\phi_{W_0}=0$ would imply $\theta=\pi/\mathfrak{a}$. As we will see in Sec. \ref{sec:EDEKKLT} and \ref{sec:EDELVS}, the choice that leads to $\theta=0$ is however important for the derivation of the EDE potential. 

With this minimisation condition we get
\begin{equation}
    V_\text{KKLT} = \frac{\mathfrak{a}^2 A^2 e^{-2 \mathfrak{a} \tau}}{6 \tau}+\frac{\mathfrak{a} A^2 e^{-2 \mathfrak{a} \tau}}{2 \tau^2}-\frac{\mathfrak{a}  A |W_0| e^{-\mathfrak{a} \tau}}{2 \tau^2} +\frac{M^2}{12 \tau^2} \,.
    \label{eq:VKKLT}
\end{equation}
This scalar potential admits a Minkowski minimum with spontaneously broken supersymmetry for $M$ and $|W_0|$ given by
\begin{equation}
\label{eqn:KKLTminima}
\begin{split}
    & |W_0| = \frac23 A\,\mathfrak{a} \tau\, e^{-\mathfrak{a}\tau} \left(1 + \frac{5}{2 \mathfrak{a} \tau}\right)\,, \\
    & M=\sqrt{2 \mathfrak{a}}  A \,e^{-\mathfrak{a}  \tau} \sqrt{\mathfrak{a}  \tau+2}\,,
    \end{split}
\end{equation}
where the gravitino mass in Planck units scales as
\begin{equation}
m_{3/2} = \frac{A\, \mathfrak{a}}{3 \sqrt{2 \tau} }\, e^{-\mathfrak{a} \tau}.
\label{gravitinoKKLT}    
\end{equation}

The minimum may be further lifted to a small but non-zero cosmological constant via a small shift in $M$. We will use (\ref{eqn:KKLTminima}) when discussing parameter values in KKLT.

\subsection{Large Volume Scenario}

We now turn to the LVS \cite{Balasubramanian:2005zx,Cicoli:2008va}. We will divide this discussion in two, each with a different choice of the underlying structure of the Calabi-Yau (CY) manifold. First, we assume a so called `Swiss cheese' CY, where, given the field content $T_b = \tau_b + i \theta_b$ and $T_s = \tau_s + i \theta_s$ (representing `big' and `small' 4-cycle volume moduli, respectively), the total volume takes the form
\begin{equation}
    \v = \tau_b^{3/2}-\tau_s^{3/2} \, .
\end{equation}
Now, adding a nilpotent superfield $X$ as in the KKLT case, we have the following K\"ahler potential
\begin{equation}
K = -2 \ln\b{\v+\frac{\hat\xi}{2}} + \frac{\bar{X}X}{\v^{2/3}} \,,
\label{KLVS}
\end{equation}
where $\mathcal{O}(\alpha'^3)$ corrections are proportional to $\hat\xi \equiv \xi\,g_s ^{-3/2}$ with $\xi = -\frac{\zeta(3)\chi}{2(2\pi)^3}$ where $\chi$ is the CY Euler number. Furthermore, in order to generate a potential for the fields, we take a superpotential with the contribution coming from non-perturbative corrections as
\begin{equation}
W = W_0 + M X + A_s\, e^{-\mathfrak{a}_s T_s} + A_b \,e^{-\mathfrak{a}_b T_b} \,.
\label{eqn:WLVS}
\end{equation}
Computing the scalar potential we find 2 contributions: a leading one responsible for stabilizing the volume, $\tau_s$ and the axion $\theta_s$ in a Minkowski (or slightly dS) vacuum, and a subleading one stabilizing the axion $\theta_b$. The scalar potential is then given by
\begin{equation}
V = V_\text{LVS}\b{\v,\tau_s,\theta_s} + V_b\b{\theta_b}\,,
\end{equation}
where, in detail, we have the LVS potential (at leading order in the $\v \gg 1$ and $\mathfrak{a}_s\tau_s\gg 1$ expansions, and setting again $W_0=-|W_0|$ and $A_s=|A_s|$)
\begin{equation}
V_\text{LVS}\b{\v,\tau_s,\theta_s} = \frac{8 \mathfrak{a}_s^2 A_s^2 e^{-2 \mathfrak{a}_s \tau_s} \sqrt{\tau_s}}{3 \v} - \frac{4 \mathfrak{a}_s A_s \tau_s |W_0|\,e^{-\mathfrak{a}_s \tau_s}}{\v^2}\cos\b{\mathfrak{a}_s \theta_s} + \frac{3|W_0|^2\hat\xi}{4\v^3} +\frac{M^2}{\v^{4/3}}\,,
\label{VLVS1}
\end{equation}
where one can immediately see that the axion gets stabilized at $\theta_s = 0$, and the potential for the axion $\theta_b$ 
\begin{equation}
\label{eqn:Vb}
    V_b\b{\theta_b} = -\frac{4 \mathfrak{a}_b A_b |W_0|\, e^{-\mathfrak{a}_b \v^{2/3}}}{\v^{4/3}}\cos\b{\mathfrak{a}_b\theta_b}\,,
\end{equation}
which fixes the axion $\theta_b$ at $\theta_b = 0$. The leading order LVS potential, at $\theta_s=0$, therefore reads 
\begin{equation}
V_{\rm LVS} = \frac{8 \mathfrak{a}_s^2 A_s^2 e^{-2 \mathfrak{a}_s \tau_s} \sqrt{\tau_s}}{3 \v} - \frac{4 \mathfrak{a}_s A_s \tau_s |W_0|\,e^{-\mathfrak{a}_s \tau_s}}{\v^2} + \frac{3|W_0|^2\hat\xi}{4\v^3}+\frac{M^2}{\v^{4/3}}\,.
\label{eqn:LVS_pot}
\end{equation}
As shown in detail in App. \ref{AppLVS}, this potential admits a global Minkowski minimum at
\begin{eqnarray}
\v &\simeq& \frac{3 |W_0| \sqrt{\tau_s}}{4 \mathfrak{a}_s A_s}\,e^{\mathfrak{a}_s \tau_s}\simeq |W_0|\,e^{\frac{\mathfrak{a}_s}{g_s}\left(\frac{\xi}{2}\right)^{2/3}}\,,\qquad 
\tau_s = \left(\frac{\xi}{2}\right)^{2/3}\frac{1}{g_s} \,,
\label{LVSmin} \\
M^2 & = & \frac{27}{20} \frac{|W_0|^2}{\mathfrak{a}_s}\,\frac{\sqrt{\tau_s}}{\v^{5/3}}\,.
\end{eqnarray}
Moreover, the gravitino mass turns out to be
\begin{equation}
m_{3/2} = \frac{|W_0|}{\v} \simeq e^{- \frac{\mathfrak{a}_s}{g_s}\left(\frac{\xi}{2}\right)^{2/3}} \,.
\label{eqn:m32LVS}
\end{equation}

We now turn to a Calabi-Yau with a K3 or $T^4$ fibration over a $\mathbb{P}^1$ base and a diagonal del Pezzo divisor \cite{Candelas:1993dm, Cicoli:2011it}. In this case the volume is written in terms of the 3 \Kahler moduli as
\begin{equation}
    \v = \sqrt{\tau_1}\tau_2-\tau_s^{3/2} \,.    
\end{equation}
This class of manifolds was used in the context of Fibre Inflation \cite{Cicoli:2008gp}, where the inflaton is the direction $u=\tau_1/\tau_2$ orthogonal to the overall volume mode. The \Kahler potential assumes the form\footnote{The actual moduli dependence of the uplifting contribution in fibred CY cases might be more complicated since it might involve both the overall volume and the fibre modulus. In this case, perturbative corrections to $K$ should be used to fix the fibre modulus in terms of the overall volume to obtain an uplifting term of the standard form.}
\begin{equation}
    K = -2 \ln\b{\v+\frac{\hat\xi}{2}}+\frac{\bar{X}X}{\v^{2/3}}\,.
\end{equation}
Furthermore, the superpotential takes the form
\begin{equation}
    W = W_0 + M X + A_s \,e^{-\mathfrak{a}_s T_s} + A_1 \,e^{-\mathfrak{a}_1 T_1}+ A_2 \,e^{-\mathfrak{a}_2 T_2} \,.
\end{equation}
After stabilizing the axion at $\theta_s = 0$, one finds again the uplifted potential given in \eqref{eqn:LVS_pot}, following the same minimisation conditions as in the Swiss cheese case. 
Given that in this case, in the $\tau_s\to 0$ limit, the volume is determined by 2 fields, rather than the 1 of the simpler Swiss cheese geometry, one direction in the $(\tau_1,\tau_2)$-plane is left flat by LVS stabilization. This flat direction can be lifted by string loops \cite{Cicoli:2008gp, Cicoli:2016xae, Cicoli:2017axo, AbdusSalam:2022krp}, higher derivative \cite{Cicoli:2016chb} or non-perturbative corrections \cite{Cicoli:2011yy, Lust:2013kt} to the action and can play a role either in early \cite{Cicoli:2008gp,Cicoli:2011ct,Broy:2015zba,Cicoli:2016chb, Cicoli:2016xae, Cicoli:2017axo, Burgess:2016owb} or late time cosmology \cite{Cicoli:2012tz}. Finally, the potential of the two bulk axions $\theta_1$ and $\theta_2$ is generated at an even subleading order by the $T_1$- and $T_2$-dependent non-perturbative contributions to the superpotential.

\section{Odd Axions and Moduli Stabilization}\label{Sec:Axions}

\subsection{Axions in string theory}

For a detailed treatment of axions in string theory, we refer the reader to \cite{Svrcek:2006yi, Conlon:2006tq, Grimm:2007hs, Cicoli:2012sz}. Here we provide a brief overview, covering the necessary background for the detailed discussion to follow.

In addition to the fundamental axion $C_0$, axions emerge in the 4-dimensional low-energy effective theory of type IIB string compactifications from dimensional reduction of $p$-form gauge fields. The shift-symmetry which earns these fields the name `axion' corresponds to gauge invariance of the higher dimensional theory, and the small axion mass, like a standard field theory axion, is generated by non-perturbative effects, such as gaugino condensation and instantons.

We define the 4-dimensional axion fields as,
\begin{equation}
    b^a = \int_{\Sigma_a} B_2 \ , \qquad   c^a =  \int_{\Sigma_a} C_2 \ ,  \qquad \theta^\alpha = \int_{D_\alpha} C_4\ ,  \label{IIBaxions}
\end{equation}
where $B_2$, $C_2$, and $C_4$, correspond to the Kalb-Ramond $2$-form and the Ramond-Ramond $2$- and $4$-form fields, respectively, and $\Sigma_a$ and $D_\alpha$ denote respectively a basis of $2$-cycles and $4$-cycles of the underlying CY three-fold $\mathcal{X}$ with $a=1,..,h^{1,1}_-(\mathcal{X})$ and $\alpha=1,..,h^{1,1}_+(\mathcal{X})$. Here $h^{1,1}_\pm(\mathcal{X})$ are the so-called Hodge numbers which count the number of holomorphic $(1,1)$-forms of $\mathcal{X}$ which are even or odd under the orientifold involution with $h^{1,1}=h^{1,1}_+ +h^{1,1}_-$. 

The $C_4$ axions are inextricably linked to stabilization of the volume moduli in both the KKLT and LVS approaches, as can be appreciated from (\ref{VKKLT}) for KKLT and (\ref{VLVS1}) and (\ref{eqn:Vb}) for LVS. In KKLT, stabilization of the volume $\tau$ necessitates stabilization of the $C_4$ axion $\theta$, and similarly in LVS stabilization of the small cycle volumes $\tau_s$ necessitates stabilization of $\theta_s$. The $\theta_b$ axion, partner to the LVS large cycle volume $\tau_b$, is ostensibly decoupled from stabilization of $\tau_b$ which is fixed by perturbative effects. 

The $B_2$ axions are also linked to the stabilization of the volume moduli due to the mixing between $b$ and $\tau$ fields in the K\"ahler potential which breaks the shift symmetry of the $B_2$ axions at the perturbative level. Consequently, any effect that stabilizes the K\"ahler moduli generates also a potential for the $b$-fields \cite{Cicoli:2021tzt}. The shift symmetry is also broken by D-branes through the DBI action, which manifests in the 4-dimensional theory as a D-term \cite{Jockers:2005zy,Haack:2006cy}. These effects generically stabilize $b$ at a high energy scale, allowing $b$ to be neglected in analyses of cosmology, as in \cite{Long:2014dta}.

The situation is instead different for the $C_2$ axions which are not a priori linked to the stabilization of the volume moduli, and therefore provide an opportunity for cosmological model building. In fact, the shift symmetry of the $C_2$ axions is unbroken by many of the standard ingredients of flux compactifications, but can be broken upon the inclusion of different effects we study in this work, such as fluxed ED1-instantons or gaugino condensation on D7-branes in the presence of worldvolume fluxes. 

Finally, the $C_0$ axion is stabilized by the superpotential induced by background $3$-form fluxes
\be
W_0 = \int \left( F_3 - S H_3 \right) \wedge \Omega\,,
\ee
where $\Omega$ is the CY holomorphic $(3,0)$-form and $S = e^{-\phi} +{\rm i}C_0$ is the axio-dilaton which set the string coupling as $e^{-\phi} \equiv s = 1/g_s$. This generates a mass for $C_0$ that is comparable to that of the dilaton, and thus stabilization of the dilaton precludes $C_0$ from playing a cosmological role (with some exceptions \cite{Blumenhagen:2014nba}).

Let us point out that axions can also be eaten up by anomalous $U(1)$s in the process of anomaly cancellation. In this case they would become as heavy as the string scale, and so would disappear from the low energy theory. Investigations in this direction \cite{Allahverdi:2014ppa} have shown that $C_4$ and $C_2$ axions are eaten up by anomalous $U(1)$s only in the presence of D3-branes at singularities. Throughout this work we will always focus on branes wrapping cycles in the geometric regime, where therefore $C_4$ and $C_2$ axions are guaranteed to survive in the 4-dimensional theory (in this case the modes eaten up correspond to open string axions).

\subsection{Odd axions in effective field theory}

Let us now focus on the description of odd axions in type IIB string theory compactified on an orientifolded Calabi-Yau manifold $\X$ \cite{Cicoli:2021tzt}. Let us denote the basis of $2$- and $4$-forms as
\begin{eqnarray} 
    \hat{D}_\alpha&\in& H^{1,1}_+(\X), \qquad \tilde{D}^\alpha\in H^{2,2}_+(\X)\,,  \qquad \alpha=1,...,h_+^{1,1}(\X),\\
    \hat{D}_a &\in& H^{1,1}_-(\X), \qquad \tilde{D}^a\in H^{2,2}_-(\X)\,,  \qquad a=1,...,h_-^{1,1}(\X)
\end{eqnarray}
which lead to the normalization and intersection numbers
\begin{eqnarray}
    \int_\X \hat{D}_\alpha\wedge\hat{D}_\beta\wedge\hat{D}_\gamma &=& k_{\alpha\beta\gamma}\,, \qquad
    \int_\X \hat{D}_\alpha\wedge\hat{D}_a\wedge\hat{D}_b= k_{\alpha a b}\,, \\
    \int_\X \hat{D}_\alpha \wedge \tilde{D}^\beta &=& \delta^\alpha_\beta\,,\qquad \int_\X \hat{D}_a \wedge \tilde{D}^b = \delta^a_b \,.
\end{eqnarray}
Furthermore, the K\"ahler form, $C_4$, $C_2$ and $B_2$ can be expanded as \cite{Grimm:2004uq}
\begin{align}
    & J = t^\alpha \hat{D}_\alpha\,, \qquad B_2 = b^a \hat{D}_a\, \qquad C_2 = c^a \hat{D}_a, \\
    &C_4 = D_2^\alpha\wedge\hat{D}_\alpha + V^K\wedge\alpha_K - V_K\wedge\beta^K    -\theta_\alpha\tilde{D}^\alpha\,,
\end{align}
where $(\alpha_K,\beta^K)\in H^3_+(\X)$ is a basis of symplectic forms such that $\int_X \alpha_K \wedge \beta^J = \delta_K^J$. These combine to give the chiral coordinates of the $\N=1$ supergravity effective theory that read
\begin{align}
    &G^a =\bar{S} b^a + \i c^a = \frac{b^a}{g_s}+\i(c^a-C_0 b^a) \,,\qquad \tau_\alpha = \frac12\,k_{\alpha\beta\gamma}\,t^\beta t^\gamma\,,\\
    &T_\alpha = \tau_\alpha + \i \theta_\alpha - \frac14\, g_s k_{\alpha a b} G^a (G+\bar{G})^b\,.
\label{ChiralCoord}
\end{align}
The CY volume is an implicit function of the $T_\alpha$ and $G^a$ fields
\begin{equation}
    \v = \frac16\, k_{\alpha\beta\gamma} t^\alpha t^\beta t^\gamma\,,
\end{equation}
which determines the tree-level K\"ahler potential
\begin{equation}
    K = -2\ln \v \,.
    \label{Ktree}
\end{equation}

We now restrict ourselves to the simple case with $h^{1,1}=2$ and $h^{1,1}_+=h^{1,1}_-=1$, where the orientifold image of the divisor $D_1$ is $D_2$. It is therefore possible to define an orientifold-even 4-cycle $D_+ \equiv  D_1 \cup D_2$ and and orientifold-odd 4-cycle $D_-\equiv D_1 \cup (-D_2)$. Hence the K\"ahler form and the K\"ahler modulus take the form
\begin{equation}
J = t\,\hat{D}_+\,,\qquad
T = \tau + i \theta -\frac14 g_s k G(G+\bar{G})\,,
\label{chiralCoordinates}
\end{equation}
where $k_{+--}\equiv k$, $\tau_+\equiv \tau$ and $\hat{D}_+$ is the $2$-form Poincaré dual to $D_+$. Defining $k_{+++}\equiv \tilde{k}$, the CY volume takes therefore the form
\begin{equation}
\mathcal{V} = \frac13\sqrt{\frac{2}{\tilde{k}}}\,\tau^{3/2}\,,
\qquad 2\tau = T+\bar{T}-\gamma \b{G+\bar{G}}^2 = 2 \, \Re(T) -\frac{4\gamma}{g_s^2}\,b^2\,,
\label{Vodd}
\end{equation}
where we have introduced
\begin{equation}
\label{eqn:gamma}
    \gamma \equiv - \frac{1}{4}g_s k\,.
\end{equation}
Substituting (\ref{Vodd}) in (\ref{Ktree}), we can obtain the K\"ahler potential which becomes
\begin{equation}
K = -3 \ln \left(\Re(T) - \frac{2\gamma}{g_s^2}\,b^2\right).
\label{NewK}
\end{equation}
Note the explicit dependence of the K\"ahler potential on the $B_2$ axion, implying that this field does not enjoy any perturbative shift symmetry. Moreover, the sign of $\gamma$ is required to be positive ($\gamma > 0\,\Leftrightarrow \, k <0$) in order to avoid the $C_2$ axion being a ghost. This can be seen by calculating the kinetic term for the orientifold odd axion 
\begin{equation}
\L_\text{kin} = K_{G\bar{G}}\,\p_\mu G\, \p^\mu \bar{G} \supset K_{G\bar{G}}\,\p_\mu c\, \p^\mu c = \frac{3\gamma}{\tau}(\p c)^2 \, ,
\end{equation}
where $K_{G\bar{G}} = \p_{G}\p_{\bar{G}} K$ and we have set $b=0$. One can then define the canonically normalized field as
\begin{equation}
\label{eqn:canon_c2}
    \varphi = \sqrt{\frac{6\gamma}{\tau}}\, c \, .
\end{equation}

\subsection{Odd axions and non-perturbative effects}
\label{NonPertSection}

Type IIB compactifications feature different non-perturbative effects that can generate corrections to both the superpotential and the K\"ahler potential. We discuss now how these effects can break the perturbative shift symmetry of $C_4$ and $C_2$ axions.

\subsubsection{ED3-instantons}
\label{sec:ED3}

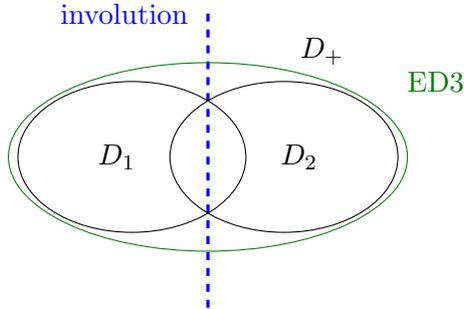
\begin{figure}[H]
\centering
\begin{tikzpicture}
     \draw[blue, dashed, very thick] (0, -2) -- (0, 2);
    \node[circle, draw=black, minimum size=2.5cm, xscale=1.2, yscale=0.8] at (-1, 0) {};
    \node[circle, draw=black, minimum size=2.5cm, xscale=1.2, yscale=0.8] at (1, 0) {};
    \node at (-1.2, 0) {$D_1$};
    \node at (1.2, 0) {$D_2$};
    \node[circle, draw=green!50!black, minimum size=5cm, xscale=1.05, yscale=0.5] at (0, 0) {};
    \node[green!50!black] at (3, 1) {ED$3$};
    \node[black] at (1.5, 1.4) {$D_+$};
    \node[blue] at (-1.1, 1.9) {involution};
\end{tikzpicture}
\caption{ED3-instanton wrapping the smooth orientifold-even divisor $D_+$.\label{fig:smoothNOD7}}
\end{figure}

A typical source of non-perturbative corrections to the superpotential is 4-cycles wrapped by Euclidean D3-brane (ED3) instantons \cite{Witten:1996bn}. The simplest configurations are fluxless but ED3-instantons can also support non-zero $2$-form fluxes as studied in \cite{Grimm:2011dj,Grimm:2007xm}. In order to obtain a non-zero contribution to $W$ the ED3 should wrap an orientifold-even rigid cycle, which in our simple case can only be $D_+$ which we assume to be a smooth and connected divisor (see Fig. \ref{fig:smoothNOD7}). Moreover, in the case of a rank-1 instanton, a non-zero $W$ is compatible only with a purely odd $2$-form flux $F_2 = \hat{\mathfrak{f}}_- \hat{D}_-$, while rank-2 instantons can contribute to $W$ also for even fluxes \cite{Berglund:2012gr}.

Restricting just to odd fluxes, the resulting contribution to the superpotential is \cite{Grimm:2011dj,Grimm:2007xm}
\begin{equation}
W_{\rm ED3} =\, \sum_{\substack{n\in\mathbb{N} \\ \hat{\mathfrak{f}}_-\in\mathbb{Z}}} A_{n,\hat{\mathfrak{f}}_-} e^{-2\pi n \b{T + k \hat{\mathfrak{f}}_- G +\frac{1}{2} k \hat{\mathfrak{f}}_-^2 \bar{S}}} \,.
\label{ED3}
\end{equation}
This expression shows that in general ED3-instantons generate a scalar potential for $C_4$ axions, while they lift $C_2$ axions only in the presence of fluxes.

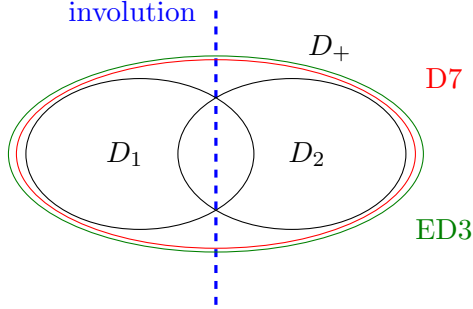
\begin{figure}[H]
\centering
\begin{tikzpicture}
    \draw[blue, dashed, very thick] (0, -2) -- (0, 2);
    \node[circle, draw=black, minimum size=2.5cm, xscale=1.2, yscale=0.8] at (-1, 0) {};
    \node[circle, draw=black, minimum size=2.5cm, xscale=1.2, yscale=0.8] at (1, 0) {};
    \node at (-1.2, 0) {$D_1$};
    \node at (1.2, 0) {$D_2$};
    \node[circle, draw=red, minimum size=5cm, xscale=1.05, yscale=0.5] at (0, 0) {};
    \node[red] at (3, 1) {D$7$};
    \node[blue] at (-1.1, 1.9) {involution};
    \node[circle, draw=green!50!black, minimum size=5.2cm, xscale=1.05, yscale=0.5] at (0, 0) {};
    \node[black] at (1.5, 1.4) {$D_+$};
    \node[green!50!black] at (3, -1) {ED$3$};
\end{tikzpicture}
\caption{ED3-instanton and a stack of D7-branes wrapping the smooth orientifold-even divisor $D_+$.\label{fig:smoothWITHD7}}
\end{figure}

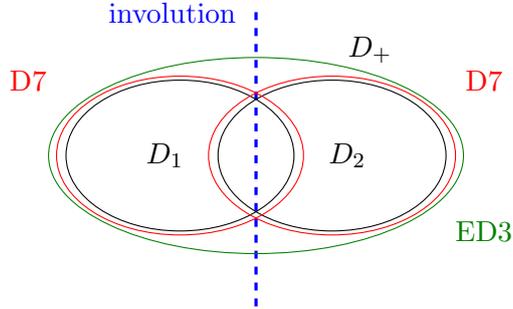
\begin{figure}[H]
\centering
\begin{tikzpicture}
      \draw[blue,dashed, very thick] (0, -2) -- (0, 2);
    \node[circle, draw=black, minimum size=2.5cm, xscale=1.2, yscale=0.8] at (-1, 0) {};
    \node[circle, draw=black, minimum size=2.5cm, xscale=1.2, yscale=0.8] at (1, 0) {};
    \node at (-1.2, 0) {$D_1$};
    \node at (1.2, 0) {$D_2$};
    \node[circle, draw=red, minimum size=2.6cm, xscale=1.25, yscale=0.81] at (1, 0) {};
    \node[circle, draw=red, minimum size=2.6cm, xscale=1.25, yscale=0.81] at (-1, 0) {};
    \node[red] at (3, 1) {D$7$};
    \node[red] at (-3, 1) {D$7$};
    \node[circle, draw=green!50!black, minimum size=5.2cm, xscale=1.05, yscale=0.5] at (0, 0) {};
    \node[black] at (1.5, 1.4) {$D_+$};
    \node[blue] at (-1.1, 1.9) {involution};
    \node[green!50!black] at (3, -1) {ED$3$};
\end{tikzpicture}
\caption{ED3-instanton wrapping the smooth orientifold-even divisor $D_+$ and a stack of D7-branes wrapping $D_1$ and its orientifold image $D_2$. \label{New}}
\end{figure}

Interestingly, in the presence of D7-branes some of the terms of this series can be absent due to gauge invariance. To see this more precisely, let us consider a stack of $N$ D7-branes wrapped either on $D_+$,  as shown in Fig. \ref{fig:smoothWITHD7}, or on $D_1$ and its orientifold image $D_2$, as shown in Fig. \ref{New}.

The $T$- and $G$-fields can get charged under the diagonal $U(1)$ of the stack of $N$ D7-branes. $G$ develops a non-zero charge $q_G$ due to a geometric St\"uckelberg mechanism only when the D7s wrap $D_1$, while the $U(1)$ charge of $T$, $q_T$, is due to the world-volume flux on the D7-stack $\hat{F}_2=\mathfrak{f}_+ \hat{D}_+ + \mathfrak{f}_- \hat{D}_-$:\footnote{More precisely, the even flux on the ED3 and the D7-stack contains also a contribution from the $B_2$-field which can be either $0$ or $1/2$, so that $\mathcal{F}_2=F_2-B_2$ for both of them. For non-spin cycles, Freed-Witten anomaly cancellation forces a half integer contribution to the even $F_2$ flux which can be cancelled by choosing $b_+=1/2$ so that the total even flux for the ED3 is zero. With this choice of $B_2$-field, the gauge flux on the D7s is then just given by $\hat{F}_2$ with integer quanta when the D7s wrap $D_+$, while a half integer contribution should be added to the even flux $\hat{F}_2$ when the D7s wrap $D_1$. In this case, we shall however omit this contribution and consider it implicitly included in $\mathfrak{f}_+$.}
\begin{eqnarray}
{\rm D7\,\, on\,\,} D_+: \quad q_G &=& 0\,, \qquad q_T = - 2 N \,\tilde{k}\, \mathfrak{f}_+ \,,  \\
{\rm D7\,\, on\,\,} D_1: \quad q_G &=&  N\,, \qquad q_T = - N \left(\tilde{k}\, \mathfrak{f}_+ + k\, \mathfrak{f}_- \right).
\end{eqnarray}
Consequently the ED3-instanton acquires a $U(1)$ charge given by \cite{Grimm:2011dj}
\begin{eqnarray}
{\rm D7\,\, on\,\,} D_+:\quad    q &=& 2\, n\, N  \,\tilde{k}\, \mathfrak{f}_+\,, 
\label{qED31} \\
{\rm D7\,\, on\,\,} D_1: \quad  q &=& n\,N  \left[k\left(\mathfrak{f}_- - \hat{\mathfrak{f}}_- \right) + \tilde{k}\, \mathfrak{f}_+\right],   
\label{qED32}
\end{eqnarray}
which in general induce non-zero charges for all terms in the ED3-instanton series (\ref{ED3}). In order to obtain a gauge invariant contribution to the superpotential, each of these terms has therefore to be multiplied by an operator of the form $\mathcal{O}\sim \Pi_i \Phi_i$ involving a product of open string modes whose $U(1)$-charge cancels the one of the instanton. However, if these are visible sector fields, they have to acquire a vanishing vacuum expectation value in order not to break the Standard Model gauge symmetry at high scales. Hence, if the Standard Model lives on the D7-stack under consideration, the only possibility to have a non-zero ED3-contribution to $W$ is by choosing the flux quanta such that $q=0$ without the need of any field-dependent prefactor. However this is never possible when the D7s wrap $D_+$ since $\tilde{k}\, \mathfrak{f}_+$ is necessarily non-zero given that it is proportional to the number of chiral states on the D7-stack. This kills any possible ED3 contribution to $W$, which is a manifestation of the known tension between chirality and moduli stabilization \cite{Blumenhagen:2007sm}. On the other hand, when the D7s wrap $D_1$, the $U(1)$-charge (\ref{qED32}) can be vanishing for an appropriate value of $\hat{\mathfrak{f}}_-$. In fact, the only non-zero contribution in the ED3-expansion (\ref{ED3}) is the one corresponding to $\mathfrak{f}_-$ such that
\begin{equation}
   \hat{\mathfrak{f}}_-= \mathfrak{f}_-   + \frac{\tilde{k}}{k}\, \mathfrak{f}_+ \equiv \mathfrak{f}\,.
\end{equation}
Thus the ED3-series (\ref{ED3}) would reduce to:
\begin{eqnarray}
{\rm SM \,\,D7\,\, on\,\,} D_+:\quad W_{\rm ED3} &=& 0 \,, \\
 {\rm SM\,\,D7\,\, on\,\,} D_1:\quad   W_{\rm ED3} &=&\, \sum_{n\in\mathbb{N}} A_{n,\mathfrak{f}}\, e^{-2\pi n \b{T + k  \mathfrak{f} G +\frac{1}{2} k \mathfrak{f}^2 \bar{S}}}  
    \label{eqn:WED3}
\end{eqnarray}
On the other hand, if the D7-stack wrapping $D_+$ or $D_1$ is a hidden sector, open string fields can acquire non-zero vacuum expectation values, and so all terms in (\ref{ED3}) can in principle survive.

\subsubsection{Gaugino condensation on D7-branes}

\begin{figure}[H]
\centering
\begin{tikzpicture}
    \draw[blue,dashed, very thick] (0, -2) -- (0, 2);
    \node[circle, draw=black, minimum size=2.5cm, xscale=1.2, yscale=0.8] at (-1, 0) {};
    \node[circle, draw=black, minimum size=2.5cm, xscale=1.2, yscale=0.8] at (1, 0) {};
    \node at (-1.2, 0) {$D_1$};
    \node at (1.2, 0) {$D_2$};
    \node[circle, draw=red, minimum size=2.6cm, xscale=1.25, yscale=0.81] at (1, 0) {};
    \node[circle, draw=red, minimum size=2.6cm, xscale=1.25, yscale=0.81] at (-1, 0) {};
    \node[red] at (3, 1) {D$7$};
    \node[red] at (-3, 1) {D$7$};
    \node[blue] at (-1.1, 1.9) {involution};
\end{tikzpicture}
\caption{A stack of D7-branes wrapped around $D_1$ and its orientifold image $D_2$.\label{fig:NONsmooth}}
\end{figure}
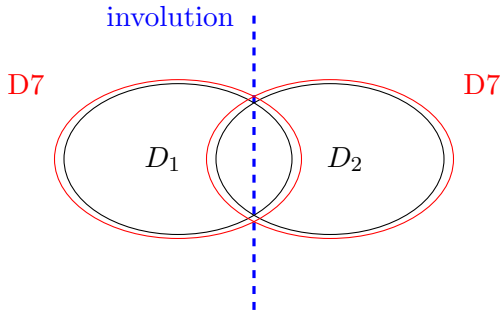

The superpotential can in general also receive a non-zero contribution from gaugino condensation in the gauge theory living on a stack of $N$ D7-branes. We shall consider the case where $N$ D7-branes wrap $D_1$ and $N$ D7-branes wrap its orientifold image $D_2$, as shown in Fig. \ref{fig:NONsmooth} (similar considerations apply for the case when the D7-branes wrap $D_+$). In this case the world-volume theory is an $SU(N)$ gauge theory and we shall allow for a general 
gauge flux of the form $F_2 = \mathfrak{f}_+ \hat{D}_+ + \mathfrak{f}_- \hat{D}_-$.

In this case the induced $W$ is given in terms of the gauge kinetic function $f_{\rm D7}$ as
\begin{equation}
W_{\rm D7}  = A \, e^{-\frac{2\pi}{N} f_{\rm D7}}  \,,
\label{eqn:WD7}
\end{equation}
where $f_{\rm D7}$ reads
\begin{eqnarray}
{\rm D7\,\, on\,\,}D_+:\quad f_{\rm D7}&=& T + k \mathfrak{f}_-\, G + \frac{1}{2} \left(k \mathfrak{f}_-^2+\tilde{k} \mathfrak{f}_+^2\right) \bar{S}\,, \label{fD71} \\
{\rm D7\,\, on\,\,}D_1:\quad f_{\rm D7}&=& T + k \left(\mathfrak{f}_+ +\mathfrak{f}_-\right) G + \frac{1}{2} \left(k \mathfrak{f}_-^2+\tilde{k} \mathfrak{f}_+^2 + 2k \mathfrak{f}_+\mathfrak{f}_-\right) \bar{S}\,.
\label{fD72}
 \end{eqnarray}
Note that $W_{\rm D7}$ would have a $U(1)$-charge of the form
\begin{eqnarray}
{\rm D7\,\, on\,\,} D_+:\quad    q &=& 2\, N  \,\tilde{k}\, \mathfrak{f}_+\,,  \\
{\rm D7\,\, on\,\,} D_1: \quad  q &=& N  \left(\tilde{k} - k\right) \mathfrak{f}_+ \,,  
\end{eqnarray}
which could vanish if $\mathfrak{f}_+=0$ (or also for $\tilde{k}=k$ when gaugino condensation is on $D_1$). For $\mathfrak{f}_+\neq 0$, the $U(1)$-charge is non-zero, and so gaugino condensation can generate an Affleck-Dine-Seiberg non-zero contribution to the superpotential \cite{Affleck:1983mk} only in the presence of a prefactor $\mathcal{O}$ which depends on chiral matter fields with appropriate $U(1)$-charges to make $W_{\rm D7}$ gauge invariant. Clearly these fields need also to develop non-zero vacuum expectation values, which is not necessarily a problem if the $SU(N)$ theory undergoing gaugino condensation belongs to a hidden sector. Let us however point out that the generation of a non-zero $W_{\rm D7}$ should be studied carefully since this situation is more complicated than the simplest one with no gauge fluxes where the world-volume theory is a pure $SU(N)$ gauge theory that is known to undergo gaugino condensation. 

Comparing (\ref{eqn:WD7}) with (\ref{eqn:WED3}) one immediately sees that, for $N>1$, ED3 contributions are subleading in respect to the one from gaugino condensation on D7-branes, and thus can be safely ignored when $W_{\rm D7}$ is generated. 

An intriguing possibility, which is clearly harder to realize explicitly, is when branes in the same stack are differently magnetized. In this case the original $SU(N)$ theory factorizes into $SU(N_1)\times SU(N_2)\times...\times SU(N_p)$ with $N_1+N_2+...N_p=N$ allowing, in principle, for multiple gaugino condensation contributions to $W$ where each of them takes the same form as (\ref{eqn:WD7}):
\begin{equation}
W_{\rm D7}  = \sum_{i=1}^p\,A_i \, e^{-\frac{2\pi}{N} f_{{\rm D7},i}} \,.
\label{eqn:WD7Gen}
\end{equation}
Alternatively, multiple non-perturbative corrections to $W$ due to gaugino condensation could arise from different stacks of D7-branes wrapped around $4$-cycles which are distinct representatives of the same homology class \cite{Long:2014dta}.

\subsubsection{ED1-instantons and gaugino condensation on D5-branes}
\label{ED1D5}

Another potential source of non-perturbative corrections to the effective action are ED1-instantons and gaugino condensation on D5-branes wrapping internal 2-cycles. Due to holomorphy, these effects are expected to correct the K\"ahler potential but not the superpotential \cite{McAllister:2008hb, Cicoli:2021tzt, Cicoli:2021gss}. This can be seen as follows. Due to the general arguments presented in \cite{Witten:1996bn}, any non-perturbative correction to the superpotential should go to zero either in the large volume limit or for vanishing string coupling. Hence ED1/D5 non-perturbative corrections to $W$ should depend on the volume of the wrapped 2-cycle $t$ as $W_{\rm ED1/D5} \sim e^{-(t+G)}$, so that $W_{\rm ED1/D5}\to 0$ for $t\to \infty$. However, as can be seen from (\ref{ChiralCoord}), $t$ is not a correct chiral coordinate for the type IIB supergravity effective theory. The correct chiral superfield is instead $T \sim t^2$. Thus, the putative superpotential $W_{\rm ED1/D5} \sim e^{-(\sqrt{T}+G)}$ would not be a holomorphic function, and so it is expected to vanish. On the other hand, note that a non-zero non-perturbative $W$ could arise from gaugino condensation on D5-branes wrapping vanishing 2-cycles \cite{Grimm:2007hs}, though this in turn introduces new subtleties, in particular, control of the effective field theory around the singularity, and that axions can be `eaten' up by anomalous $U(1)$'s at singularities \cite{Cicoli:2012vw,Cicoli:2013mpa,Cicoli:2013cha}.

We shall therefore ignore potential ED1/D5 corrections to $W$ but we will consider the possibility of non-perturbative corrections to $K$ since this quantity is not protected by holomorphy. These corrections have not been computed explicitly in type IIB (see however \cite{Camara:2008zk} for a derivation of in type I toroidal orbifolds). However the authors of \cite{McAllister:2008hb, Cicoli:2021gss} estimated the scaling of the leading ED1/D5 corrections to the K\"ahler potential. Here, we generalize their results proposing an educated guess for the series of non-perturbative corrections to $K$ from ED1/D5-branes wrapped on an internal 2-cycle $t$ with non-zero odd gauge flux. Making an analogy with the ED3 case (\ref{ED3}), we propose
\begin{equation}
K_{\rm ED1} = -3 \ln \left(\Re(T) - \frac{2\gamma}{g_s^2}\,b^2 + ...+\sum_{\substack{n\in\mathbb{N} \\ \hat{\mathfrak{f}}_-\in\mathbb{Z}}} A_{n,\hat{\mathfrak{f}}_-} e^{-2\pi n \left(\frac{t}{\sqrt{g_s}} + k \hat{\mathfrak{f}}_- G \right)}\right),
\label{ED1}
\end{equation}
where 
\begin{equation}
t = \sqrt{\frac{2}{\tilde{k}}\left(\Re(T) - \frac{2\gamma}{g_s^2}\,b^2\right)}\,,
\end{equation}
and the dots denote perturbative corrections in $\alpha'$ and $g_s$, as well as non-perturbative worldsheet $\alpha'$ corrections \cite{Grimm:2007xm}, which do not depend on the $C_2$ axion due to its shift symmetry. In (\ref{ED1}) we have absorbed in the prefactor $A_{n,\hat{\mathfrak{f}}_-}$ a potential dilaton-dependent factor $e^{-2\pi n\frac12 k \hat{\mathfrak{f}}_-^2 \bar{S}}$. 

Making an analogy with the D7 case (\ref{eqn:WD7Gen}), we can also propose a similar form for the corrections to the K\"ahler potential for the case of multiple gaugino condensates on a stack of D5-branes (absorbing again in the prefactors potential $\bar{S}$-dependent exponents)
\begin{equation}
K_{\rm D5} = -3 \ln \left(\Re(T) - \frac{2\gamma}{g_s^2}\,b^2 + ...+\sum_{i=1}^p A_i\, e^{-\frac{2\pi}{N_i} \left(\frac{t}{\sqrt{g_s}} + k \mathfrak{f}_i G \right)}\right).
\label{D5}
\end{equation}

\subsubsection{ED(-1)-instantons and gaugino condensation on D3-branes}

Other possible non-perturbative effects in type IIB can be generated by ED(-1)-instantons or gaugino condensation on D3-branes at singularities. We shall however not consider these corrections to the superpotential since they depends just on the dilaton (and the blow-up mode resolving the local singularity $T_{\rm loc}$) \cite{Cicoli:2012fh}, $W_{\rm ED(-1)/D3} \sim e^{-(S+T_{\rm loc})}$. Moreover, in this case, $C_4$ and $C_2$ axions tend to be removed from the low energy theory since they get eaten up by anomalous $U(1)$s localized at the singularity \cite{Allahverdi:2014ppa}.

\subsection{Odd axions and D-terms}
\label{sec:Dterms}

As we have explained above, the shift symmetry of $C_2$ axions can be broken only in the presence of non-zero $2$-form fluxes. However these world-volume fluxes generate also moduli-dependent Fayet-Iliopoulos (FI) terms for the diagonal $U(1)$ of D7-branes. We need therefore to analyze these FI-terms carefully. 

\subsubsection{Fayet-Iliopoulos terms}

For a stack of D7-branes wrapping the divisor $D_{\rm D7}$ with world-volume flux $\mathcal{F}_2= F_2-B_2$ the FI-term takes the form \cite{Jockers:2005zy,Haack:2006cy}
\begin{equation}
\xi_{\rm FI} = \frac{1}{4\pi\v} \int_{D_{\rm D7}} J\wedge \mathcal{F}_2\,.
\label{FI}
\end{equation}
The K\"ahler form can be expanded as $J=t^\alpha \hat{D}_\alpha$, while the gauge flux $\mathcal{F}$ can be decomposed as (without including potential half integer contributions for the even $B_2$ field)
\begin{equation}
\mathcal{F}_2= F_2-B_2 = \mathfrak{f}^\alpha \hat{D}_\alpha +\left(\mathfrak{f}^a-b^a\right) \hat{D}_a\,.
\end{equation}
Focusing for concreteness on the simple case with $h^{1,1}=2$ and $h^{1,1}_+=h^{1,1}_-=1$, the exact expression of the FI-term (\ref{FI}) depends on the nature of the divisor $D_{\rm D7}$ wrapped by the D7-branes. If $D_{\rm D7}=D_+$, as in Fig. \ref{fig:smoothWITHD7}, we have
\begin{equation}
\xi_{\rm FI} = \frac{1}{4\pi\v} \int_{D_+} J\wedge \mathcal{F}_2= \left(\frac{3\tilde{k}}{4\pi}\right) \frac{\mathfrak{f}_+}{\tau}\,,
\label{FIplus}
\end{equation}
which does not depend on the $B_2$ axion since $k_{++-}=0$. Note that for and ED3-instanton (\ref{FIplus}) is identically zero since the flux can only be purely odd, i.e. $\mathfrak{f}_+=0$, to have a non-zero contribution to $W$.

On the other hand, if $D_{\rm D7}=D_1$, as in Fig. \ref{New}, the FI-term (\ref{FI}) takes the form
\begin{eqnarray}
\xi_{\rm FI} &=& \frac{1}{4\pi\v} \int_{D_1} J\wedge \mathcal{F}_2 
= \frac{1}{8\pi\v} \int_{D_+} J\wedge \mathcal{F}_2 +
\frac{1}{8\pi\v} \int_{D_-} J\wedge \mathcal{F}_2 \nonumber \\
&=& \frac{3}{8\pi\tau}\left[\tilde{k}\mathfrak{f}_+ + k \left(\mathfrak{f}_- -b\right)\right],
\label{FI1}
\end{eqnarray}
which introduces an explicit dependence on the $B_2$ axion, even for $F_2=0$. In what follows we shall assume, without loss of generality, $F_2 = \mathfrak{f} \hat{D}_1 - \mathfrak{f} \hat{D}_2 = \mathfrak{f} \hat{D}_-$, which implies $\mathfrak{f}_-=\mathfrak{f}$ and $\mathfrak{f}_+=0$. This choice guarantees that ED3/D7 non-perturbative corrections to $W$ maintain their dependence on the $G$-field but simplifies the expression of the FI-term (\ref{FI1}) to
\begin{equation}
\xi_{\rm FI} = \left(\frac{3k}{8\pi}\right) \frac{\left(\mathfrak{f} -b\right)}{\tau}\,.
\label{FI1new}
\end{equation}

The total D-term potential includes also open string modes $\chi_i$ with $U(1)$ charges $q_i$ and looks like
\begin{equation}
V_D = \frac{g^2}{2}\left(\sum_i q_i|\chi_i|^2 - \xi_{\rm FI}\right)^2
\quad\text{with}\quad g^2 = \frac{4\pi}{{\rm Re}(T)}\,.
\label{VDtot}
\end{equation}
The D-term potential scales as $V_D\sim \xi_{\rm FI}^2/\tau \sim \mathcal{V}^{-2}$, and so it is leading with respect to the F-term potential due to the no-scale cancellation. Hence, the minimum is located at
\begin{equation}
q\, |\chi|^2 \simeq \xi_{\rm FI}\,,
\label{Dstab}
\end{equation}
where for simplicity we have focused just on a single charged matter field whose charge $q$ has an opposite sign with respect to the FI-term. All charged matter fields whose charge has the same sign as $\xi_{\rm FI}$ are instead fixed to zero. 

As can be seen from (\ref{FI1new}), $\xi_{\rm FI}$ is in general a function of two fields: $\tau$ and $b$. Hence, the relation (\ref{Dstab}) fixes one direction, among $|\phi|$, $\tau$ and $b$, in terms of the other two which, at this level of approximation, remain still flat. They are lifted by subdominant F-term contributions. Before studying the stabilization of these two directions, let us point out that the axionic partner of the saxionic direction fixed by the D-terms is eaten up by the anomalous $U(1)$ via the St\"uckelberg mechanism. This axion is in general a linear combination of the axionic phase $\zeta$ of $\phi=|\phi|\,e^{i \zeta}$, the $C_4$ axion $\theta$, and the $C_2$ axion $c$. The resulting $U(1)$ mass is proportional to the decay constants of these axions \cite{Arkani-Hamed:1998ufq}
\begin{equation}
M_{U(1)}^2 \sim g^2 \left(f_\zeta^2 + f_\theta^2 + f_c^2\right),
\end{equation}
where 
\begin{equation}
f_\zeta^2 \simeq \xi_{\rm FI}\sim \frac{(\mathfrak{f}-b)}{\tau}\,,\qquad f_\theta^2 \simeq K_{T\bar{T}}\sim \frac{1}{\tau^2}\,,\qquad f_c^2 \simeq K_{G\bar{G}} \sim \frac{g_s}{\tau}\,.
\label{fs}
\end{equation}
In the presence of a hierarchy among these decay constants, the combination of axions eaten up by the anomalous $U(1)$ is mostly given by the axion with the largest decay constant. We will see that F-term stabilization gives two branches for $b$:
\begin{itemize}
\item $b=0$: in this case the Abelian gauge boson becomes massive by eating up the open string axion $\zeta$ since
\begin{equation}
f_\zeta^2 \sim \frac{\mathfrak{f}}{\tau}\gg f_c^2 \sim \frac{g_s}{\tau} \gg f_\theta^2 \sim\frac{1}{\tau^2} \quad \text{for} \quad  \tau^{-1} \lesssim \mathcal{O}(0.01) \ll g_s \sim \mathcal{O}(0.1) \ll \mathfrak{f}\sim \mathcal{O}(1)\,.
\end{equation}

\item $b=\mathfrak{f}$: in this case the Abelian gauge boson becomes massive by eating up the $C_2$ axion $c$ since
\begin{equation}
f_c^2 \sim \frac{g_s}{\tau} \gg f_\theta^2 \sim\frac{1}{\tau^2} \gg f_\zeta^2 \sim 0 \quad \text{for} \quad  \tau^{-1} \lesssim \mathcal{O}(0.01) \ll g_s \sim \mathcal{O}(0.1) \,.
\end{equation}
\end{itemize}

\subsubsection{$B_2$ axion stabilization}
\label{B2stab}

The two directions left flat by D-term stabilization are lifted by F-term contributions. The matter fields receive F-term contributions from soft supersymmetry breaking terms of order the gravitino mass 
\begin{equation}
V_F(|\chi|) = \mathcal{C}\,m_{3/2}^2\, |\chi|^2 +... = \frac{|W_0|^2}{\mathcal{V}^2}\, |\chi|^2 +... \,,
\label{VFchi}
\end{equation}
where $\mathcal{C}$ is an $\mathcal{O}(1)$ coefficient and the dots denote potential contributions with higher powers of $|\chi|$. On the other hand, the F-term potential for the K\"ahler moduli $\tau$ and $b$ is given in KKLT by (\ref{eq:VKKLT}) and in LVS by (\ref{eqn:LVS_pot}) after including the dependence on $b$ through the mixing in the K\"ahler coordinates, as can be seen from (\ref{Vodd}). Using the general formalism developed in \cite{Cicoli:2021tzt}, in the KKLT case we obtain (after fixing the $C_4$ axion)
\begin{equation}
V_\text{KKLT}(\tau,b) \simeq \frac{\mathfrak{a}^2 A^2 e^{-2 \mathfrak{a} \left(\tau + \tilde{\gamma}\, b^2\right)}}{6 \tau}-\frac{\mathfrak{a}  A |W_0| e^{-\mathfrak{a} \left(\tau + \tilde{\gamma}\, b^2\right)}}{2 \tau^2} +\frac{M^2}{12 \tau^2} \,,
\label{eq:VKKLTNew}
\end{equation}
where we have defined $\tilde{\gamma}\equiv 2\gamma/g_s^2$ and we have included only the leading terms for $\mathfrak{a}\tau \gg 1$ and $\tau \gg \tilde{\gamma} b^2$.

In LVS we will consider only the case where the $G$-modulus mixes with the big modulus $T_b$, i.e. $k_{s--}=0$ while $k_{b--}\neq 0$, since the huge $e^{-\mathfrak{a}_b\tau_b}\lll 1$ suppression is crucial to reproduce the correct EDE scale. Hence the total LVS potential including the $B_2$ axion becomes
\begin{equation}
V_{\rm LVS} (\mathcal{V},\tau_s, \theta_b, b) = V_{\rm LVS} (\mathcal{V},\tau_s) -\frac{4 \mathfrak{a}_b A_b |W_0|\, e^{-\mathfrak{a}_b \left(\tau_b + \tilde{\gamma}\, b^2\right)}}{\v^{4/3}}\cos\b{\mathfrak{a}_b\theta_b}\,,
\label{eqn:LVS_potNew}
\end{equation}
where $V_{\rm LVS} (\mathcal{V},\tau_s)$ is given by (\ref{eqn:LVS_pot}) with the axion $\theta_s$ fixed at zero, and we have included again only the leading terms for $\tau_b \gg \tilde{\gamma} b^2$.

Let us analyze the KKLT and LVS cases separately for D7-branes wrapping either $D_+$ of $D_1$.
\begin{itemize}
\item \textbf{KKLT with D7s on $D_+$ or $D_1$:}
D-term fixing gives
\begin{equation}
q\,|\chi|^2 \sim \frac{\left(\mathfrak{f}-d\,b\right)}{\tau}\,,
\label{FIset}
\end{equation}
where $d=0$ and $\mathfrak{f} = \mathfrak{f}_+$ when $D_{\rm D7}=D_+$, while $d=1$ when $D_{\rm D7}=D_1$. The relation (\ref{FIset}) fixes $|\chi|$ in terms of $\tau$ and $b$ (or just $\tau$ for $D_{\rm D7}=D_+$). The remaining flat directions are fixed by $V_{\rm KKLT}(\tau, b)$ since this potential dominates over $V_F(|\chi|)$. In fact, substituting (\ref{FIset}) in (\ref{VFchi}) we obtain
\begin{equation}
V_F(|\chi|) \sim \left(\mathfrak{f}-d\,b\right) \frac{|W_0|^2 }{\tau^4} \ll V_{\rm KKLT} \sim \frac{|W_0|^2}{\tau^3}\quad\text{for}\quad \tau\gg 1\,.
\end{equation}
It is then straightforward to realize that $V_{\rm KKLT}$ fixes $b=0$ and $\tau$ as in KKLT, implying that the axion eaten up by the anomalous $U(1)$ is $\zeta$. 

\item \textbf{LVS with D7s on $D_+$:}
D-term moduli stabilization sets
\begin{equation}
q\,|\chi|^2 \sim \frac{\mathfrak{f}_+}{\mathcal{V}^{2/3}}\,,
\end{equation}
which fixes $|\chi|$ in terms of $\v$. Substituting this result in (\ref{VFchi}) we obtain
\begin{equation}
V_F(|\chi|) \sim \mathfrak{f}_+ \frac{|W_0|^2 }{\v^{8/3}}\,,
\label{VupT}
\end{equation}
that represents the standard expression for T-brane dS uplifting \cite{Cicoli:2015ylx}. Hence, in this case the UV consistency of the underlying model forces the presence of a precise dS uplifting source, in addition to the potential existence of anti D3-branes. The K\"ahler moduli and the $B_2$ axion are then stabilized by (\ref{VupT}) together with the LVS potential (\ref{eqn:LVS_potNew}) which fix $b=0$ and the $T$-moduli as in standard LVS construction. This implies that the axion eaten up by the anomalous $U(1)$ is again the open string mode $\zeta$.

\item \textbf{LVS with D7s on $D_1$:}
D-term fixing implies
\begin{equation}
q\,|\chi|^2 \sim \frac{\left(\mathfrak{f}-b\right)}{\mathcal{V}^{2/3}}\,,
\label{FIsetNew}
\end{equation}
which fixes $b$ in terms of $|\chi|$ and $\v$ that have to be considered as two independent variables. Given that the LVS potential (\ref{eqn:LVS_potNew}) does not depend on $|\chi|$, the charged matter field has to be stabilized by its F-term potential (\ref{VFchi}). Two different situations can arise:
\begin{enumerate}
\item If $\mathcal{C}>0$, the minimum for $|\chi|$ lies at $|\chi|=0$. Substituting this result back in (\ref{FIsetNew}) we find $b=\mathfrak{f}$ which implies that the $C_2$ axion is removed from the low energy effective theory since it is eaten up by the anomalous $U(1)$. The location of the $\v$ minimum remains instead the same as in the LVS case without odd moduli since it is still determined by the leading order potential $V_{\rm LVS}$ in (\ref{eqn:LVS_potNew}).

\item If $\mathcal{C}<0$, the matter field is tachyonic and can develop a non-zero vacuum expectation value. For example, if the F-term potential for $|\chi|$ features an additional cubic contribution of the form (setting $|W_0|\sim \mathcal{O}(1)$):
\begin{equation}
V_F(|\chi|) = - \frac{1}{\mathcal{V}^2}\, |\chi|^2 + \frac{1}{\mathcal{V}^\alpha}\, |\chi|^3 \,,
\label{VFchiNew}
\end{equation}
with $\alpha>0$, the minimum of the charged matter field is located at $|\chi|\sim \v^{\alpha-2}$. Substituting this relation back in (\ref{FIsetNew}) we obtain
\begin{equation}
\left(\mathfrak{f}-b\right) \sim \v^{2\left(\alpha-\frac53\right)}\,.
\end{equation}
For $\alpha<5/3$, $\left(\mathfrak{f}-b\right)$ is $\v$-suppressed, and so the solution is given again, at first approximation, by $b\simeq f$, implying that the $C_2$ axion is eaten up. On the other hand, for $\alpha \geq 5/3$, $\left(\mathfrak{f}-b\right)$ becomes larger than unity. Consequently, as can be seen from (\ref{fs}), the axion eaten up by the anomalous $U(1)$ becomes the open string mode $\zeta$ since its decay constant becomes larger than the one of the $C_2$ axion. As in the previous case, the $T$-moduli are still fixed by the standard LVS potential.
\end{enumerate}
\end{itemize}

The best case scenario is therefore the one where the $B_2$ axion is fixed at zero, so that the $C_2$ axion can survive in the low energy effective theory and play the role of the EDE field. Let us finally stress that, at this level of approximation, the $B_2$ axion becomes massive, while the $C_2$ axion is still massless. If the potential for the $C_2$ axion is generated by subleading non-perturbative effects, the resulting moduli mass spectrum would feature a hierarchical structure with the lightest mode given by the $C_2$ axion.

\section{EDE in KKLT}
\label{sec:EDEKKLT}

As our first string theory realization of EDE, we consider the KKLT scenario. In this case the EDE field has to be a $C_2$ axion since $C_4$ axions would be too heavy given that in KKLT the moduli are stabilized by non-perturbative effects, and so $C_4$ axions are as heavy as the K\"ahler moduli. 

We therefore focus on $C_2$ axions and build our model following the recipe given in \cite{McDonough:2022pku} to ensure the correct shape of the EDE potential. The K\"ahler potential and superpotential are given by
\begin{eqnarray}
K &=& - 3 \ln \left[ T + \bar{T} -  \gamma(G+ \bar{G})^2\right] +3\,\frac{\bar{X}X}{T+\bar{T}}\,, 
\label{KKKLTC2} \\
W &=& W_0 + M X + A\, e^{- \mathfrak{a} T}  + A_1 \,e^{- \tilde{\mathfrak{a}}(T+ k \mathfrak{f}_1\, G)} +  A_2\, e^{- \tilde{\mathfrak{a}}(T+k\mathfrak{f}_2\, G)} + A_3 \, e^{- \tilde{\mathfrak{a}}(T+k\mathfrak{f}_3\, G)}\,,
\label{WKKLTC2}
\end{eqnarray}
with
\begin{equation}
\mathfrak{a}=\frac{2\pi}{N}<\tilde{\mathfrak{a}}=\frac{2\pi}{M}
\qquad\Leftrightarrow\qquad M<N\,,
\label{aCond}
\end{equation}
to ensure that the EDE scale, $\sim$ eV, is naturally suppressed with respect to the standard KKLT potential. This exponential suppression removes the EDE fine-tuning to obtain the correct EDE scale previously argued in \cite{McDonough:2022pku} since the parameters $A$ and $A_i$ ($i=1,2,3$) can take natural $\mathcal{O}(1)$ values in Planck units. Moreover, in order to generate the desired periodicity of the EDE potential, we have to impose
\begin{equation}
\mathfrak{f}_1=\mathfrak{f}\,,\qquad \mathfrak{f}_2 = 2\mathfrak{f}\,, \qquad \mathfrak{f}_3 = 3 \mathfrak{f} \,.
\label{FluxCond}
\end{equation}

According to our discussion of non-perturbative effects in type IIB compactifications presented in Sec. \ref{NonPertSection}, this situation can be reproduced at the microscopic level in two different possible ways:
\begin{enumerate}
\item \textbf{ED3-instantons on $D_+$:} the superpotential (\ref{WKKLTC2}) can arise from fluxed ED3-instantons wrapped around $D_+$ if this divisor does not intersect with a D7-stack supporting the Standard Model. For $\mathfrak{a}=2\pi n$ and $\tilde{\mathfrak{a}}=2\pi m$, the condition (\ref{aCond}) can be met if $m>n$. On the other hand, the condition (\ref{FluxCond}) can be satisfied if different terms in the instanton expansion compete among each other. Clearly, the underlying assumption is that all the other terms in the expansion are either absent or suppressed.

\item \textbf{Gaugino condensation on D7-branes:} as we have seen, in the presence of gaugino condensation on D7-branes, ED3-instanton contributions are subdominant and can be safely neglected. In this case, the superpotential (\ref{WKKLTC2}) can be reproduced by gaugino condensation on 4 stacks of D7-branes wrapping the same cycle. This can arise, for example, if there are 4 distinct representatives of the same homology class or if different branes of the same stack are differently magnetized. 1 D7-stack is not fluxed and consists of $N$ D7-branes with $\mathfrak{a} = 2\pi/N$. On the other hand, the other 3 stacks have the same number of D7-branes $M<N$, so that $\tilde{\mathfrak{a}} =2\pi/M$ and the condition (\ref{aCond}) is met. Moreover, these 3 D7-stacks should carry different world-volume fluxes to satisfy the periodicity condition (\ref{FluxCond}).
\end{enumerate}

Note that, as explained in Sec. \ref{NonPertSection}, each of the 3 fluxed non-perturbative effects in (\ref{WKKLTC2}) should come along with extra dilaton-dependent exponential suppressions of the form:
\begin{equation}
A_i = \tilde{A}_i\,e^{-\tilde{\mathfrak{a}}\, \mathfrak{f}_i^2 \bar{S}/2}\ll 1 \qquad \forall\,i=1,2,3\,.
\label{Apref}
\end{equation}
If present, these exponential suppression factors would require some fine tuning to obtain the correct $\left[1-\cos\left(\varphi/f\right)\right]^3$ shape of the EDE potential. In fact, the condition (\ref{FluxCond}), when inserted in (\ref{Apref}), implies that the 3 prefactors $A_i$ are not all of the same order if all $\tilde{A}_i$ are $\mathcal{O}(1)$ coefficients, implying the need to tune the $\tilde{A}_i$ appropriately. 

We shall however exploit model building to avoid this tuning by noticing that (\ref{fD71}) and (\ref{fD72}) allow for a cancellation of the $\bar{S}$-dependent part of the gauge kinetic function if even fluxes are turned on. This is always possible for gaugino condensation on D7-branes and ED3-instantons with rank $m>1$ \cite{Berglund:2012gr}, which is actually forced to be the case due to the $m>n\geq 1$ condition. When the ED3/D7-stack is wrapping $D_+$, if $\tilde{k} = - p^2 k$ with $p \in \mathbb{N}$ (where in particular $p\neq 0$), (\ref{fD71}) reduces to 
\begin{equation}
f_{\rm D7} = T + k \mathfrak{f}_-\, G + \frac{k}{2} \left( \mathfrak{f}_-^2 - p^2 \mathfrak{f}_+^2\right) \bar{S} = T + k \mathfrak{f}_-\, G\,,
 \end{equation}
if the fluxes are chosen such that $\mathfrak{f}_- = \pm p\,\mathfrak{f}_+$. Similar considerations apply to the case when the ED3/D7-stack is wrapping $D_1$. 

We shall therefore focus on the effective field theory defined by (\ref{KKKLTC2}) and (\ref{WKKLTC2}). The resulting scalar potential can be separated into terms that scale with $e^{-\mathfrak{a}\tau}$ and a series of corrections suppressed by powers of $e^{-\tilde{\mathfrak{a}}\tau}$. We expand the scalar potential in small $e^{-\tilde{\mathfrak{a}}\tau}$, to arrive at
\begin{equation}
V = V_{\rm KKLT} + V_{\rm EDE}\,,
\end{equation}
where $V_{\rm KKLT}$ is the standard KKLT potential defined in \eqref{eq:VKKLT} and\footnote{Here and it what follows we will not include in $V_{\rm EDE}$ the constant term in (\ref{eq:EDE_V}). This contribution can be obtained by an appropriate tuning of the uplifting contribution.}
\begin{equation}
V_{\rm EDE}=\tilde{V}_0 \left[A_1 \cos (\tilde{\mathfrak{a}}|k| \mathfrak{f}\, c)+A_2 \cos (2 \tilde{\mathfrak{a}}|k| \mathfrak{f}\, c)+A_3 \cos (3 \tilde{\mathfrak{a}}|k| \mathfrak{f}\, c)\,\right].   
\label{EDEPot}
\end{equation}
The potential (\ref{EDEPot}) can reproduce the $\left[1-\cos\left(\varphi/f\right)\right]^3$ EDE potential if $A_1 = 15\, \tilde{A}/4$, $A_2 =-3\,\tilde{A}/2$ and $A_3 =\tilde{A}/4$. The EDE scale is then given by
\begin{equation}
V_0 \equiv - \tilde{A}\,\tilde{V}_0 = \frac{A \,\tilde{A}\,\left(2\tilde{\mathfrak{a}}-3\mathfrak{a}\right)\, e^{ -(\mathfrak{a}+\tilde{\mathfrak{a}})\tau}}{6 \tau^2}\,.
\label{eqn:KKLT_EDE_scale}
\end{equation}
Let us point out that a $\cos (2 \tilde{\mathfrak{a}}|k| \mathfrak{f}\, c)$ term would also arise from the mixed term between the 2 non-perturbative contributions in (\ref{WKKLTC2}) proportional to $A_1$ and $A_3$, suggesting that the potential (\ref{EDEPot}) could also be generated for $A_2=0$. However, this is not the case since it can be proven that, under the condition that the EDE field is hierarchically lighter than the K\"ahler modulus $\tau$, this mixed term has always to be negligible. Hence $A_2\neq 0$ is indeed needed to generate the $\cos (2 \mathfrak{a}|k| \mathfrak{f}\, c)$ term in the EDE potential (\ref{EDEPot}) and keep the correct hierarchy of scales.

Notice that we are computing the EDE potential at the Minkowski minimum of KKLT, namely enforcing conditions \eqref{eqn:KKLTminima}, and that, at leading order, we have the stabilization $b=g_s {\rm Re}(G)=0$ (which implies $c={\rm Im}(G)$) and $\theta ={\rm Im} (T) = 0$. As discussed in Sec.~\ref{B2stab}, when $b$ is stabilized at zero, the $C_2$ axion is not eaten up by an anomalous $U(1)$. 

Furthermore, recalling the canonical normalization for the $C_2$ axion given in \eqref{eqn:canon_c2}, we obtain a term in the potential of the form
\begin{equation}
\cos\b{\tilde{\mathfrak{a}}|k| \mathfrak{f}\, c} = \cos\b{\tilde{\mathfrak{a}} |k|\mathfrak{f} \sqrt{\frac{\tau}{6\gamma}}\,\varphi} \equiv \cos\b{\frac{\varphi}{f}}\,,
\end{equation}
giving the following decay constant
\begin{equation}
f \equiv \sqrt{\frac{6\,\gamma}{\tau}}\,\frac{1}{\tilde{\mathfrak{a}}|k|\mathfrak{f}} = \sqrt{\frac{3 g_s}{2 |k|\,\tau}} \frac{1}{\tilde{\mathfrak{a}} \mathfrak{f}} \,.
\label{decayconstantST}
\end{equation}
Upon switching to the canonically normalized EDE field $\varphi$, we may write
\begin{equation}
V_{\rm EDE} = V_0 \left[ -\frac{15}{4} \cos \b{\frac{\varphi}{f}} + \frac32 \cos\b{\frac{2\varphi}{f}} -\frac14\cos\b{\frac{3\varphi}{f}} \right] \,,
\label{KKLTfinalVEDE}
\end{equation}
where the overall scale $V_0$ can be expressed in terms of the decay constant $f$ and the gravitino mass $m_{3/2}$ as (reinstating powers of $M_P$)
\begin{equation}
V_0 = \frac{N\,\tilde{A}}{\sqrt{2}\,\tau^{3/2}} \left(\frac{2}{M}-\frac{3}{N}\right)\left(\frac{m_{3/2}}{M_P}\right)\,e^{ -\frac{3}{4\pi}
\frac{g_s M}{|k| \mathfrak{f}^2}\left(\frac{M_P}{f}\right)^2}\,M_P^4\,. 
\label{V0tuned}
\end{equation}
Note that, contrary to general expectations from the weak gravity conjecture, $V_0$ is exponentially suppressed in terms of $g_s M \left(M_P/f\right)^2$, instead of just $\left(M_P/f\right)$. For $M\gg 1$ and $f<M_P$, this helps to suppress the EDE scale and to reduce the required fine tuning on $\tilde{A}$, even if the presence of the small factor $g_s\ll 1$ does not allow to remove the tuning completely. Setting $f=0.2\,M_P$ and $\mathfrak{f}=|k|=1$, the EDE scale $V_0$ scales as
\begin{equation}
V_0 \simeq \tilde{A}\left(\frac{m_{3/2}}{M_P}\right)\,e^{ -\frac{75}{4\pi}\,g_s M}\,M_P^4\,.
\label{V0finKKLT}
\end{equation}
This relation depends on the gravitino mass. In KKLT models this is related to the mass of the K\"ahler modulus $m_\tau \simeq m_{3/2}\ln\left(M_P/m_{3/2}\right)$ which has to be above $\mathcal{O}(50)$ TeV in order to avoid any cosmological moduli problem. This implies $m_{3/2}\gtrsim \mathcal{O}(1)$ TeV. Imposing therefore a gravitino mass at the TeV-scale to maximize the suppression in (\ref{V0finKKLT}) (larger values of $m_{3/2}$ would also require a larger value of $N$), $V_0\simeq 10^{-108}\,M_P^4$ gives the value of $M$ for a given string coupling. In turn, (\ref{decayconstantST}) yields the value of $\tau$ at the minimum that, when substituted in (\ref{gravitinoKKLT}), sets the value of $N$ for natural $\mathcal{O}(1)$ values of $A$.

\begin{table}[h!]
\centering
\begin{tabular}{c|c||c|c|c|c|c|c}
$g_s$& $\tilde{A}$ & $M$ & $N$ & $\tau$ & $V_0\,10^{108}\,M_P^{-4}$ & $m_{3/2}$ (TeV) & $m_\tau$ (TeV) \\
\hline
\hline
0.1 & 1  & 340 & 3200  & 10980.7 & 1.4 & 4.6 & 155.5 \\
\hline
0.3 & 1  & 114 & 1000  & 3703.4 & 2.0 & 4.6 & 155.2 \\
\hline
0.3 & $10^{-11}$  & 100 & 750  & 2849.7 & 1.6 & 3.8 & 129.6 \\
\hline
0.3 & $10^{-27}$ & 80 & 470 & 1823.8 & 0.9 & 4.6 & 154.8 \\
\hline
0.3 & $10^{-61}$  & 36 & 85 & 369.3 & 2.2 & 3.0 & 104.0 \\
\hline
\end{tabular}
\caption{Benchmark parameters that realize the EDE potential (\ref{eq:EDE_V}) with $f=0.2\, M_P$, $|A|=1$ and $m_\tau=m_{3/2}\ln\left(M_P/m_{3/2}\right)$. Recall that $\mathfrak{a}=2\pi/N$, $\tilde{\mathfrak{a}}=2\pi/M$, with $M<N$.}
\label{C2KKLTExamples}
\end{table}

We present in Tab. \ref{C2KKLTExamples} some selected numerical examples for the present model for various choices of $g_s$, $\tilde{A}$, $M$ and $N$, all of which give rise to the correct EDE scale, decay constant and a gravitino mass at the TeV-scale. If the string coupling is kept in the regime where perturbation theory does not break down, i.e. $g_s\lesssim 0.3$, natural $\mathcal{O}(1)$ values of $\tilde{A}$ correlate with $N \sim \mathcal{O}(1000) \gg M\sim \mathcal{O}(100) \gg 1$ and larger values of $\tau$. Such large values of the ranks of the condensing gauge groups are very likely to be incompatible with D7 tadpole cancellation and to induce an uncontrolled backreaction on the internal geometry (see \cite{Louis:2012nb} for a study of the maximal rank of condensing gauge groups as a function of $h^{1,1}$ for F-theory compactifications). On the other hand, tuned values of the overall prefactor of the EDE potential of order $\tilde{A}\sim\mathcal{O}(10^{-50})$, can allow for viable models with acceptably smaller numbers of D7-branes, $N\sim \mathcal{O}(100) \gg M\sim \mathcal{O}(10) \gg 1$ and smaller values of $\tau$. Note that, at fixed $g_s$, larger values of $m_{3/2}$, as can be seen from (\ref{V0finKKLT}), would require larger values of $M$, and so from (\ref{decayconstantST}) larger values of $\tau$, which imply even larger values of $N$, as can be seen from (\ref{gravitinoKKLT}). Hence, cases with $m_{3/2}$ considerably above the TeV-scale are highly disfavored. Let also point out that in general ED3-instantons  would not give the required EDE decay constant  since viable models with $\tau\gg 1$ require large values of $M$.

Summarizing, our analysis shows that EDE can be realized in KKLT with a $C_2$ axion whose potential is generated by gaugino condensates on D7-branes with world-volume fluxes. When the string coupling is small enough to trust the string loop expansion and $m_{3/2}\gtrsim \mathcal{O}(1)$ TeV, matching the EDE scale seems to require a substantial tuning of the prefactors of these non-perturbative effects. If instead $\tilde{A}$ takes natural $\mathcal{O}(1)$ numbers, then the number of D7-branes becomes too large to be compatible with a controlled effective field theory. The best scenarios correlate with a TeV-scale gravitino mass.

\section{EDE in the Large Volume Scenario}
\label{sec:EDELVS}

We will now turn our attention to another class of models, built using the LVS. We will analyze the possibility of building the EDE potential from $C_2$ axions, as in the previous KKLT construction, and the new option of using $C_4$ axions. This last option is possible in LVS but not in KKLT. In fact, in LVS models the big cycle $\tau_b$ is much heavier than the corresponding axion $\theta_b$ since $\tau_b$ is stabilized by perturbative $\alpha'$ effects which do not lift $\theta_b$.

\subsection{EDE from $C_4$ axions}
\label{sec:EDELVSC4}

Consider a `Swiss cheese' manifold with a large 4-cycle with size $\tau_b$ and a small 4-cycle with size $\tau_s$. The low energy supergravity action is defined by the following K\"ahler potential and superpotential:
\begin{eqnarray}
K &=& - 2 \ln\b{\tau_b^{3/2} -  \tau_s ^{3/2}+\frac{\hat\xi}{2}}+\frac{\bar{X}X}{\v^{2/3}} \,, \\
W &=& W_0 + M X + A_s \,e^{- \mathfrak{a}_s T_s}  + A_1\, e^{- \mathfrak{a}_1 T_b} +  A_2\, e^{- \mathfrak{a}_2 T_b} + A_3\, e^{- \mathfrak{a}_3 T_b}\,.
\end{eqnarray} 
As explained in Sec. \ref{NonPertSection}, the non-perturbative corrections to $W$ could arise from either gaugino condensation on D7-branes or ED3-instantons, where in this case we are focusing on situations without orientifold-odd moduli and vanishing world-volume fluxes (more precisely, in the case of ED3-instantons, flux-dependent contributions would be exponentially suppressed in the dilaton with respect to the leading fluxless term). Moreover, in order to engineer the desired EDE periodic potential, we proceed similarly to (\ref{FluxCond}) and require
\begin{equation}
\mathfrak{a}_1=\mathfrak{a}_b\,\qquad   \mathfrak{a}_2=2\mathfrak{a}_b\,, \qquad \mathfrak{a}_3=3\mathfrak{a}_b\,.
\end{equation}
The scalar potential turns out to be
\begin{equation}
    V = V_\text{LVS} + V_\text{EDE}\,,
\end{equation}
with $V_\text{LVS}$ as in \eqref{eqn:LVS_pot} and the EDE part given by
\begin{equation}
\label{eq:VEDELVSC4fiber}
    V_\text{EDE} = V_0 \left[\tilde{A}_1\,\cos\b{\mathfrak{a}_b \theta_b}+\tilde{A}_2\, e^{-\mathfrak{a}_b \tau_b}\cos\b{2 \mathfrak{a}_b \theta_b}+\tilde{A}_3\, e^{-2 \mathfrak{a}_b \tau_b}\cos\b{3 \mathfrak{a}_b \theta_b}\right]\,,
\end{equation}
where the EDE scale reads (for $\tau_b\simeq \v^{2/3}$)
\begin{equation}
V_0 = \frac{4 \mathfrak{a}_b\,|W_0|}{\v^{4/3}}\,A_b\,e^{-\mathfrak{a}_b \tau_b}\,,
\label{V0C4LVS}
\end{equation}
and we have redefined $A_i \equiv A_b\tilde{A}_i$ ($i=1,2,3$) to factor out in $V_0$ and overall coefficient $A_b$. This example displays a crucial difference with respect to the KKLT $C_2$ example of Sec.~\ref{sec:EDEKKLT}: the moduli dependence, namely on $\tau_b$, cannot be included completely into the overall normalization $V_0$. Instead, the 3 periodic terms appear in $V_{\rm EDE}$ with different powers of $e^{- \mathfrak{a}_b \tau_b}$, which must be compensated with an exponential hierarchy in $\tilde{A}_i$ ($i=1,2,3$) if one is to recover \eqref{eq:EDE_V} for the EDE potential.\footnote{The different scaling of each term with $\tau_b$ would also backreact on the vacuum expectation value of $\tau_b$, even if this effect is tiny since $V_{\rm EDE}$ is hierarchically smaller than $V_{\rm LVS}$.} This situation is also different from the model of \cite{McDonough:2022pku}, where each harmonic has the same modulus-dependent suppression, which can in fact be reabsorbed into $V_0$.

This model requires also an exponential tuning of the overall prefactor $A_b$ in order to match the EDE scale since $C_4$ axions do not give rise to any violation of the weak gravity conjecture (see Introduction). To see this more in detail, let us compute the EDE decay constant. The kinetic terms for the EDE field look like
\begin{equation}
\L_\text{kin} = K_{T_b\bar{T}_b} \p_\mu T_b \p^\mu \bar{T}_b \supset  K_{T_b\bar{T}_b} \p_\mu \theta_b \p^\mu \theta_b = \frac{3}{4\tau_b^2} \b{\p \theta_b}^2 \, ,
\end{equation}
and thus we canonically normalize the field as
\begin{equation}
    \varphi = \sqrt{\frac{3}{2}}\frac{\theta_b}{\tau_b} \, .
\end{equation}
We then obtain in the potential the term
\begin{equation}
    \cos\b{\mathfrak{a}_b\theta_b} = \cos\b{\sqrt{\frac23}\mathfrak{a}_b\tau_b\,\varphi} \equiv \cos\b{\frac{\varphi}{f}} \, ,
\end{equation}
from which we can easily see that the decay constant of the EDE field $\varphi$ is
\begin{equation}
f = \sqrt{\frac{3}{2}}\frac{1}{\mathfrak{a}_b\tau_b}\simeq 0.2\,\frac{N_b}{\tau_b}\qquad\text{for}\qquad \mathfrak{a}_b= \frac{2\pi}{N_b}\,.
\label{fC4LVS}
\end{equation}
Setting $f\simeq 0.2\,M_P$, (\ref{fC4LVS}) clearly implies $\tau_b\simeq N_b$. Given that the $\alpha'$ expansion is controlled by $\v^{-1/3} = \tau_b^{-1/2} \ll 1$, the big modulus should be at least $\tau_b \gtrsim \mathcal{O}(100)$ which requires a large number of D7-branes $N_b\gtrsim \mathcal{O}(100)$. Moreover, the EDE scale (\ref{V0C4LVS}) can be rewritten down as (reinstating appropriate powers of $M_P$)
\begin{equation}
V_0 = \frac{64 \pi^3}{3}\left(\frac{|W_0|\,A_b}{N_b^3}\right)\left(\frac{f}{M_P}\right)^2 M_P^4\,e^{-\sqrt{\frac32}\,\frac{M_P}{f}}\,,
\label{V0C4LVSnew}
\end{equation}
 which for $f\simeq 0.2\,M_P$ reduces to
\begin{equation}
V_0 \simeq  0.06 \left(\frac{|W_0|\,A_b}{N_b^3}\right) M_P^4\,.
\end{equation}
From this expression it is clear that $V_0 \sim 10^{-108}\,M_P^4$ can be achieved only by fine-tuning $A_b$ to exponentially small values since the flux superpotential $|W_0|$ cannot be taken too small, otherwise the volume modulus would become lighter than $\mathcal{O}(50)$ TeV. In fact, the volume modulus mass scales as
\begin{equation}
m_\v \simeq \frac{|W_0|\,M_P}{\tau_b^{9/4}} \gtrsim 50\,{\rm TeV}  
\qquad \Leftrightarrow\qquad |W_0|\gtrsim 2\times 10^{-14}\,\tau_b^{9/4}\,.
\label{CMPbound}
\end{equation}
For $\tau_b\gtrsim 100$, this gives also a lower bound on the gravitino mass of order:
\begin{equation}
m_{3/2} = \frac{|W_0|}{\tau_b^{3/2}}\,M_P\gtrsim 2\times 10^{-14}\,\tau_b^{3/4}\,M_P\gtrsim 1.5 \times 10^6\,{\rm GeV}\,.
\label{GravitinoBoundLVS}
\end{equation}
In Tab. \ref{C4KKLTExamples} we show two benchmark examples for $N_b=100$ and $N_b=1000$, which give $f\simeq 0.2\,M_P$ and the right EDE scale for the smallest possible value of $|W_0|$. At fixed $N_b$ and $\tau_b$, larger values of $|W_0|$ would give larger moduli masses and would require smaller $A_b$ and larger $N_s$, as can be seen from (\ref{LVSmin}).

\begin{table}[h!]
\centering
\begin{tabular}{c|c|c||c|c|c |c}
$N_b$ & $N_s$ & $\tau_b$ & $|W_0|$ & $A_b$ & $A_s$ & $V_0\,10^{108}\,M_P^{-4}$ \\
\hline
\hline
100 & 3 & 97.5  &  $6.0\times 10^{-10}$  & $5\times 10^{-92}$ & 0.29 & 1.8 \\
\hline
1000 & 4 & 974.6  &  $1.1\times 10^{-7}$  & $2\times 10^{-91}$ & 0.28 & 1.3 \\
\hline
\end{tabular}
\caption{Benchmark parameters for LVS EDE with $C_4$ axions. Each parameter set gives $f=0.2\,M_P$ and $m_\v\simeq 50$ TeV, and features $\mathfrak{a}_s=2\pi/N_s$, $\tau_s=10$ for $g_s=0.1$ and $\xi=2$.}
\label{C4KKLTExamples}
\end{table}

The need to perform a double fine-tuning on the 3 prefactors of the $T_b$-dependent non-perturbative effects, to get both the right EDE scale and periodicity, suggests that the $C_4$ axion associated with the volume modulus in LVS is not an optimal candidate for building an EDE model in string theory.

This conclusion continues to hold when we add more complication to the geometry, e.g. by considering a Calabi-Yau manifold with a fibered structure, as we will do in the following. Consider a manifold with volume given by
\begin{equation}
    \vol = \sqrt{\tau_1}\tau_2-\tau_s^{3/2} \,,
\end{equation}
and focus on the case $\tau_s \simeq \tau_1 \ll \tau_2$ such that the volume is predominantly set by $\tau_2$. The scalar potential derived from
\begin{eqnarray}
K &=& - 2 \ln\b{\vol+\frac{\hat\xi}{2}} + \frac{\bar{X}X}{\v^{2/3}} \\
W &=& W_0 + M X + A_s \,e^{- \mathfrak{a}_s T_s}  + A_1 \,e^{- \mathfrak{a}_1 T_1} +  
A_2\, e^{- 2\mathfrak{a}_1 T_1} + A_3\, e^{- 3\mathfrak{a}_1 T_1}\,,
\end{eqnarray}
would take again the form $V = V_\text{LVS} + V_\text{EDE}$ with
\begin{equation}
V_\text{EDE} = V_0 \sb{A_1\,\cos\b{\mathfrak{a}_1 \theta_1}+A_2\, e^{-\mathfrak{a}_1 \tau_1}\cos\b{2 \mathfrak{a}_1 \theta_1}+A_3\, e^{-2 \mathfrak{a}_1 \tau_1}\cos\b{3 \mathfrak{a}_1 \theta_1}} \, ,
\end{equation}
where the EDE scale scales as (for $A_i= \mathcal{A}_1 \tilde{A}_i$)
\begin{equation}
V_0 = \frac{4 \mathfrak{a}_1 \tau_1 |W_0|}{\vol^2 }\,\mathcal{A}_1\, e^{-\mathfrak{a}_1 \tau_1} \,.
\label{V0fibre}
\end{equation}
Just like in the previous case, this scenario exhibits an explicit dependence of $V_{\rm EDE}$ on 4-cycle moduli, in this case $\tau_1$, requiring an exponential hierarchy between $\tilde{A}_1$, $\tilde{A}_2$ and $\tilde{A}_3$.

However, the fibered model provides one advantage, in the form of increased flexibility in setting the EDE decay constant. Following the same procedure as before, we compute the kinetic terms at leading order in $1/\v$
\begin{equation}
    \L_{\rm kin} = K_{T_1 \bar{T}_1} \p_{\mu} T_1 \p^{\mu}\bar{T}_1 \supset K_{T_1 \bar{T}_1} \p_{\mu} \theta_1 \p^{\mu}\theta_1 =  \frac{1}{4\tau_1^2} \b{\p\theta_1}^2\,.
\end{equation}
Thus, canonically normalizing as
\begin{equation}
    \varphi = \frac{1}{\sqrt{2}\tau_1} \theta_1\,,
\end{equation}
we have that the potential will contain terms like
\begin{equation}
    \cos\b{\mathfrak{a}_1\theta_1} = \cos\b{\sqrt{2}\mathfrak{a}_1\tau_1\varphi} \equiv\cos\b{\frac{\varphi}{f}}\,,
\end{equation}
finding
\begin{equation}
f = \frac{1}{\sqrt{2}\mathfrak{a}_1\tau_1}\simeq 0.1\,\frac{N_1}{\tau_1}\qquad \text{for}\qquad \mathfrak{a}_1=\frac{2\pi}{N_1}\,,
\end{equation}
which depends only on $\tau_1$ and not $\tau_2$. Given that in anisotropic compactifications with $\tau_2\gg \tau_1$, the overall internal volume is controlled mainly by $\tau_2$, an $\alpha'$ expansion under control can be compatible with $\tau_1\sim \mathcal{O}(10)$ which removes, in turn, the need to go to a large number of D7-branes $N_1$ to reproduce $f\simeq 0.2\,M_P$. Clearly, in this case, more natural values $N_1\simeq \mathcal{O}(10)$ can be allowed. However, the system still needs a very large tuning of the prefactor $\mathcal{A}_1$ in (\ref{V0fibre}) since the EDE scale can be rewritten as (showing explicit powers of $M_P$)
\begin{equation}
V_0 = \frac{4}{\sqrt{2}|W_0|^{1/3}}\,\mathcal{A}_1\left(\frac{m_\v}{M_P}\right)^{4/3}\left(\frac{M_P}{f}\right) M_P^4\, 
e^{-\frac{1}{\sqrt{2}}\,\frac{M_P}{f}} \,,
\label{V0fibrenew}
\end{equation}
which, setting $f\simeq 0.2\,M_P$ and taking in (\ref{CMPbound}) $m_\v\simeq 50$ TeV, reduces to
\begin{equation}
V_0 = 2.4\times 10^{-19}\,\frac{\mathcal{A}_1}{|W_0|^{1/3}}\,M_P^4\,.
\end{equation}
For $|W_0|\sim\mathcal{O}(1)$, clearly $V_0 \sim 10^{-108}\,M_P^4$ requires to tune $\mathcal{A}_1$ down to $\mathcal{A}_1\sim \mathcal{O}(10^{-90})$.

The lesson to learn from these attempts of building EDE potentials by means of non-perturbative effects in $W$ is twofold: ($i$) models, where the real part of the chiral superfield used for EDE is stabilized at zero, require less tuning of the underlying parameters to reproduce the correct periodicity of the EDE potential; ($ii$) matching the EDE scale without fine-tuning any prefactor of the non-perturbative effects, which generate the EDE potential, requires a violation of the weak gravity conjecture. This singles out $C_2$ axions since they can violate the weak gravity conjecture and belong to the chiral superfield $G=\bar{S} b+{\rm i}\,c$ where the $B_2$ axion is fixed at $b=0$. With this in mind let us explore $C_2$ models in the framework of LVS.

\subsection{EDE from $C_2$ axions}
\label{sec:EDELVSC2}

We now return to the $C_2$ axion case, studied previously in the context of KKLT in Sec.~\ref{sec:EDEKKLT}. We consider two possibilities for generating the EDE potential, namely gaugino condensation on D7-branes (or fluxed ED3-instantons) and gaugino condensation on D5-branes (or fluxed ED1-instantons).

\subsubsection{Gaugino condensation on D7-branes}
\label{sec:EDELVSD7}

We focus on a situation with $h^{1,1}_+=2$ and $h^{1,1}_-=1$ where the orientifold even moduli describe a typical Swiss-cheese Calabi-Yau, while the orientifold odd modulus mixes just with the big modulus. The volume form therefore looks like
\begin{equation}
\mathcal{V}=\tau_b^{3/2}-\tau_s^{3/2}= \frac{1}{2\sqrt{2}}\left[\left(T_b+\bar{T}_b - \gamma (G + \bar{G})^2\right)^{3/2} - \left(T_s+\bar{T}_s\right)^{3/2}\right]\,.
\end{equation}
The low-energy effective action is determined by the following K\"ahler potential and superpotential
\begin{eqnarray}
K &=& -2\ln\b{\v+\frac{\hat{\xi}}{2}}+\frac{\bar{X}X}{\v^{2/3}}\,, \\
W &=&W_{\rm LVS }+A_1\, e^{-\tilde{\mathfrak{a}} (T_b+k\mathfrak{f}_1 G)}+A_2\, e^{-\tilde{\mathfrak{a}}(T_b+k\mathfrak{f}_2 G)}+A_3\, e^{-\tilde{\mathfrak{a}} (T_b+k\mathfrak{f}_3 G)}\,,
\label{WC2LVS}
\end{eqnarray}
with
\begin{equation}
    W_{\rm LVS}= W_0+ M X + A_s \,e^{-\mathfrak{a}_s T_s}+ A_b \,e^{-\mathfrak{a}_b T_b}\,.
\end{equation}
Similarly to (\ref{FluxCond}), we also impose 
\begin{equation}
\mathfrak{f}_1=\mathfrak{f}\,,\qquad \mathfrak{f}_2=2\mathfrak{f}\,,\qquad \mathfrak{f}_3= 3\mathfrak{f}\,,
\end{equation}
in order to match the periodicity of the EDE potential. $W_{\rm LVS}$ is the standard LVS superpotential with the inclusion of the $T_b$-dependent non-perturbative effect which stabilizes the $C_4$ axion $\theta_b$. The last 3 terms in (\ref{WC2LVS}) are instead responsible for the generation of the EDE potential. As explained in Sec. \ref{sec:EDEKKLT}, these can arise from branes wrapping the big divisor which can be either fluxed ED3-instantons or D7-branes with non-zero world-volume fluxes which support gaugino condensation. Similarly to Sec. \ref{sec:EDEKKLT}, we will see that ED3-instantons cannot reproduce the correct EDE decay constant for $\v\gg 1$, as well as the correct EDE scale without fine-tuning the prefactors $A_i$ ($i=1,2,3$) to exponentially small values. Moreover, when both effects are present, ED3-instantons are always subdominant with respect to gaugino condensation. In the following, we shall therefore focus mainly on gaugino condensation on D7-branes, keeping in mind however that the EDE potential could also be realized by ED3-instantons (at the price of introducing fine-tuning and working at small internal volume) if gaugino condensation effects are not generated. 

As explained in Sec. \ref{sec:EDEKKLT}, the last 3 non-perturbative effects in (\ref{WC2LVS}) receive also $\bar{S}$-dependent contributions in the exponents, which might destroy the required periodicity of the EDE potential if the corresponding prefactors take natural $\mathcal{O}(1)$ numbers. However, we have seen that these dilaton-dependent contributions can be cancelled by an appropriate choice of even and odd fluxes. In the LVS case, there is another intriguing possibility if initially the number of odd moduli is $h^{1,1}_-=2$. Note that this is not possible in KKLT since in cases where the orientifold involution exchanges two non-identical divisors \cite{Gao:2013pra, Altman:2021pyc}, $0\leq h^{1,1}_- \leq r$ for $h^{1,1}=2r$ or $h^{1,1}=2r+1$ with $r\in \mathbb{N}$, implying that $h^{1,1}_+=1$ is incompatible with $h^{1,1}_-=2$. On the other hand, LVS models with $h^{1,1}_+=2$ can feature $h^{1,1}_-=2$. In this case, if the initial divisors are $D_i$ ($i=1,2,3,4$) and the involution exchanges $D_1\leftrightarrow D_2$ and $D_3\leftrightarrow D_4$, the even and odd divisors are
\begin{eqnarray}
D_+^{(b)} = D_1 \cup D_2\,,\qquad D_-^{(\tilde{b})} = D_1 \cup (-D_2)\,, \\
D_+^{(s)} = D_3 \cup D_4\,,\qquad D_-^{(\tilde{s})} = D_3 \cup (-D_4)\,,
\end{eqnarray}
where we have assumed to have a big and a small orientifold-even modulus. For 2 orientifold-odd moduli the expression (\ref{fD72}) for the gauge kinetic function of a D7-stack wrapping $D_1$ generalizes to \cite{Grimm:2011dj} (focusing for simplicity just on the case with odd world-volume fluxes)
\begin{eqnarray}
f_{\rm D7} &=& T_b + \left(k_{b\tilde{b}\tilde{b}}\mathfrak{f}_{\tilde{b}} + k_{b\tilde{b}\tilde{s}}\mathfrak{f}_{\tilde{s}}\right) G_{\tilde{b}} +  \left(k_{b\tilde{b}\tilde{s}}\mathfrak{f}_{\tilde{b}} + k_{b\tilde{s}\tilde{s}}\mathfrak{f}_{\tilde{s}}\right) G_{\tilde{s}} \nonumber \\
&+& \frac12 \left[\left(k_{b\tilde{b}\tilde{b}}\mathfrak{f}_{\tilde{b}} + k_{b\tilde{b}\tilde{s}}\mathfrak{f}_{\tilde{s}}\right) \mathfrak{f}_{\tilde{b}} +
\left(k_{b\tilde{b}\tilde{s}}\mathfrak{f}_{\tilde{b}} + k_{b\tilde{s}\tilde{s}}\mathfrak{f}_{\tilde{s}}\right)\mathfrak{f}_{\tilde{s}}\right] \bar{S}\,,
 \end{eqnarray}
where the indices with a tilde denote odd-moduli. If $k_{b\tilde{s}\tilde{s}}=0$ and the flux quantum $\mathfrak{f}_{\tilde{b}}$ is set to zero, this expression simplifies to
\begin{equation}
f_{\rm D7} = T_b + k_{b\tilde{b}\tilde{s}}\mathfrak{f}_{\tilde{s}}\,G_{\tilde{b}}\,,  
\end{equation}
implying that the superpotential would not depend on $G_{\tilde{s}}$. The $B_2$-axion $b_{\tilde{s}}$ would however appear in the FI-term since (\ref{FI1}) would generalize to
\begin{equation}
\xi_{\rm FI} \sim \frac{t_b}{\mathcal{V}}\left[k_{b\tilde{b}\tilde{b}} b_{\tilde{b}} + k_{b\tilde{b}\tilde{s}} \left( b_{\tilde{s}} - \mathfrak{f}_{\tilde{s}}\right)\right].
\end{equation}
If $k_{b\tilde{b}\tilde{b}}=0$, the FI-term would simply depend on $b_{\tilde{s}}$ and, as explained in Sec. \ref{B2stab}, D-term stabilization would fix $b_{\tilde{s}}=\mathfrak{f}_{\tilde{s}}$ if the charged matter fields do not acquire tachyonic masses from supersymmetry breaking. This implies that the $G_{\tilde{s}}$ axion is eaten up and disappears from the effective field theory. The $B_2$ axion $b_{\tilde{b}}$ could instead be fixed at zero by subleading effects.

The total scalar potential can be written as
\be
V=V_{\rm LVS}(\v, \tau_s,\theta_s)+V_b(\theta_b, b)+V_{\rm EDE}(c)\,,
\ee
where in each term we have written down explicitly just the dependence on the moduli which get frozen by each type of contribution. $V_{\rm LVS} $ is the uplifted LVS potential (\ref{eqn:LVS_pot}) which stabilizes $\mathcal{V}$, $\tau_s$ and the axion $\theta_s$, $V_b$ is the contribution included in (\ref{eqn:LVS_potNew}) which fixes $\theta_b$ and $b=0$, while the EDE potential reads
\begin{equation}
V_{\rm EDE}= V_0\left\{ -\frac{15}{4}  \cos\left[\tilde{\mathfrak{a}} \left(k\mathfrak{f} c+ \theta_b\right)\right]+ \frac32 \cos\left[\tilde{\mathfrak{a}} \left(2 k\mathfrak{f} c+ \theta_b\right)\right] 
- \frac14 \cos\left[\tilde{\mathfrak{a}} \left(3k\mathfrak{f} c+ \theta_b\right)\right] \right\}\,,
\label{eq:VEDEaux}
\end{equation}
with
\begin{equation}
V_0 = \frac{4|W_0|\tilde{\mathfrak{a}}}{ \mathcal{V}^{4/3}}\,\tilde{A}
\,e^{- \tilde{\mathfrak{a}} \tau_b}.
\label{V0again}
\end{equation}
where we have set $A_1 = -15 \tilde{A}/4$, $A_2 = 3\tilde{A}/2$ and $A_3=-\tilde{A}/4$. In order to obtain the correct EDE potential we require $\mathfrak{a}_b<\tilde{\mathfrak{a}}$, so that $V_{\rm EDE} \ll V_b$ and the stabilization of $\theta_b$ is completely determined by $V_b$ for $b=0$. The overall sign of (\ref{eqn:Vb}) plays an important role in this EDE realization: if $A_b >0$ the bulk $C_4$ axion is stabilised at $\theta_b=\pi/\mathfrak{a}_b$, whereas if $A_b <0$ the minimum is at  $\theta_b=0$. Given the $\theta_b$ dependence of (\ref{eq:VEDEaux}) it is evident that only $A_b <0$ can lead to the desired EDE potential\footnote{More in general, if $\theta_b = \pi/\mathfrak{a}_b$, the correct EDE potential could still be obtained if $\tilde{\mathfrak{a}} = p \mathfrak{a}_b$ with $p\in \mathbb{N}$.} 
\begin{equation}
V_{\rm EDE}=V_0\left[-\frac{15}{4}\,\cos(\tilde{\mathfrak{a}} |k|\mathfrak{f}\, c)+\frac32\,\cos(2 \tilde{\mathfrak{a}} |k|\mathfrak{f}\, c)-\frac14\,\cos(3 \tilde{\mathfrak{a}} |k|\mathfrak{f}\, c)\right].
\label{VEDEC2LVS}
\end{equation}
Let us now determine the EDE decay constant. Using the canonically normalized field defined in \eqref{eqn:canon_c2}, the cosine terms in the EDE potential behave as
\begin{equation}
\cos\b{\tilde{\mathfrak{a}} |k| \mathfrak{f}\, c}=\cos\b{\tilde{\mathfrak{a}} |k| \mathfrak{f}\,\sqrt{\frac{\tau_b}{6\gamma}}\,\varphi}\equiv \cos\b{\frac{\varphi}{f}}\,,
\end{equation}
finding
\begin{equation}
f= \frac{1}{\tilde{\mathfrak{a}} |k| \mathfrak{f}}\,\sqrt{\frac{6\gamma}{\tau_b}} = \sqrt{\frac{3 g_s}{8\pi^2 |k|\mathfrak{f}^2 }}\,\frac{M}{\sqrt{\tau_b}} \qquad\text{for}\qquad \tilde{\mathfrak{a}}=\frac{2\pi}{M}\,.
\label{fLVSC2}
\end{equation}
The overall EDE scale (\ref{V0again}) therefore becomes (reinstating powers of $M_P$)
\begin{equation}
V_0 = \frac{16 (2\pi)^5 |k|^2 \mathfrak{f}^4}{9 g_s^2 M^5} |W_0|\,\tilde{A} \left(\frac{f}{M_P}\right)^4\,e^{ -\frac{3}{4\pi}
\frac{g_s M}{|k| \mathfrak{f}^2}\left(\frac{M_P}{f}\right)^2}\,M_P^4\,. 
\end{equation}
Note that the exponential suppression is the same as in (\ref{V0tuned}) since we are again using $C_2$ axions whose potential is generated by gaugino condensation on D7-branes with non-zero fluxes. Contrary to the KKLT EDE case discussed in Sec. \ref{sec:EDEKKLT}, however this case can realize EDE without the need to go to an excessively large number of D7-branes. The main reason is that in LVS, as can be seen from the minimization relations (\ref{LVSmin}), large $\tau_b$ does not require a very large number of D7-branes to avoid ultralight moduli that would induce cosmological problems. Let us see this crucial point more in detail. For $f=0.2\,M_P$ and $\mathfrak{f}=|k|=1$, the EDE scale $V_0$ reduces to
\begin{equation}
V_0 \simeq \frac{27.85 }{g_s^2 M^5}\, |W_0|\,\tilde{A} \,e^{ -\frac{75}{4\pi}\,g_s M}\,M_P^4\,,
\label{V0fin}
\end{equation}
where the lowest possible value of $|W_0|$ that maximizes the suppression is given by (\ref{CMPbound}) in terms of $\tau_b$ which is fixed by (\ref{fC4LVS}) for a given $M$ and $g_s$.  

\begin{table}[h!]
\centering
\begin{tabular}{c|c||c|c|c|c|c|c|c|c|c}
$g_s$& $\tilde{A}$ & $N_s$ & $M$ & $|W_0|$ & $\v=\tau_b^{3/2}$ & $\tau_s$ & $V_0\,10^{108}\,M_P^{-4}$ & $A_s$ & $\xi$ & $m_\v$ (TeV) \\
\hline
\hline
0.3 & 1  & 1 & 128 & 1 & $3.2\times 10^5$ & 2.37 & 2.6 & 1.70 & 1.2 & $3.2\times 10^7$ \\
\hline
0.3 & 1  & 1/2 & 121 & $2.8\times 10^{-6}$  & $2.7\times 10^5$ & 2.24 & 2.7 & 1.51 & 1.1 & 50 \\
\hline
0.1 & 1  & 2 & 362 & $3.3\times 10^{-5}$  & $1.4\times 10^6$ & 8.25 & 2.2 & 2.96 & 1.5 & 50 \\
\hline
\end{tabular}
\caption{Benchmark parameters that realize the EDE potential (\ref{VEDEC2LVS}) with $f=0.2\, M_P$. We have used the LVS minimization relations (\ref{LVSmin}) with $\mathfrak{a}_s=2\pi/N_s$ and $\tilde{\mathfrak{a}}=2\pi/M$. The case with $N_s=1/2$ corresponds to a rank-2 ED3-instanton \cite{Berglund:2012gr}.}
\label{C2LVSExamples}
\end{table}

We present in Tab. \ref{C2LVSExamples} three numerical examples with a different value of $g_s$ which reproduce the correct EDE scale without the need to tune the prefactor $\tilde{A}$. Contrary to the KKLT case discussed in Sec. \ref{sec:EDEKKLT}, there is no need to have $\mathcal{O}(1000)$ D7-branes. If $\tilde{A}$ is kept of order unity, the number of D7-branes $M$ has to be $M\sim \mathcal{O}(100)$ which is however realizable in F-theory compactifications \cite{Louis:2012nb}. Smaller values of $M$ would require an exponentially small $\tilde{A}$ and would also reduce the value of $\tau_b$ due to the need to reproduce $f\simeq 0.2\,M_P$ from (\ref{fLVSC2}). The first case in Tab. \ref{C2LVSExamples} is the most generic since the flux superpotential takes the natural value $|W_0|=1$ which correlates with $m_\v \sim \mathcal{O}(10^{10})$ GeV and $m_{3/2} \sim \mathcal{O}(10^{13})$ GeV. On the other hand, the last two cases in Tab. \ref{C2LVSExamples} are characterized by lower moduli masses, $m_\v \simeq 50$ TeV and $m_{3/2} \simeq 10^6$ GeV, due to the tuning of $|W_0|$ to small values. Let us stress that in all cases the CY volume is large enough to trust the effective field theory.

To complete our analysis, let us consider also K3-fibered LVS compactifications with volume $\v=\sqrt{\tau_1}\tau_2-\tau_s^{3/2}$ since, as we have already seen in Sec. \ref{sec:EDELVSC4}, they give more freedom in matching the EDE energy scale and decay constant if $\v$ is anisotropic with $\tau_2 \gg \tau_1\gg 1$. If the $G$-modulus mixes only with $T_1$, the superpotential would still be given by (\ref{WC2LVS}) but with the substitution $T_b \to T_1$. The stabilisation of $\theta_1$ would proceed as the stabilization of $\theta_b$ above, while $\theta_2$ would in practice remain as a massless spectator field. The EDE decay constant would still be given by (\ref{fLVSC2}) but again with the substitution $\tau_b\to \tau_1$. The EDE potential would take the same form as in (\ref{eq:VEDEaux}) but the EDE scale would become (for $f=0.2\,M_P$ and $\mathfrak{f}=|k|=1$)
\begin{equation}
V_0 \simeq 23.9\, \frac{|W_0|}{\v^2} \,g_s\,M\,\tilde{A} \,e^{ -\frac{75}{4\pi}\,g_s M}\,M_P^4\,.
\label{V0finFibred}
\end{equation}
The difference with the previous case is that in fibered CY models, the lightest modulus is the direction $u$ orthogonal to the volume mode which is stabilized beyond leading LVS order. Imposing that its mass is above the bound from the cosmological moduli problem, we find \cite{Cicoli:2018cgu}
\begin{equation}
m_u \simeq \frac{|W_0|}{\v^{3/2}\tau_1^{1/4}}\,M_P \gtrsim 50\,{\rm TeV}  
\qquad \Leftrightarrow\qquad \v \lesssim 1.3\times 10^9\,|W_0|^{2/3}\,\tau_1^{-1/6}\,.
\end{equation}
When $\tau_1$ is fixed around $\tau_1\sim\mathcal{O}(10^4)$ by the requirement to obtain $f\simeq 0.2\,M_P$, this condition can clearly be compatible with $|W_0|\sim\mathcal{O}(1)$ since it would just require $\v \lesssim \mathcal{O}(10^8)$. Setting $\v \sim \mathcal{O}(10^8)$ would indeed correspond to an anisotropic extra-dimensional volume with $\tau_2\sim\mathcal{O}(10^6)\gg \tau_1\sim\mathcal{O}(10^4)\gg 1$. We present in Tab. \ref{C2LVSExamplesNew} two numerical examples with a different value of $g_s$ which reproduce the correct EDE scale without the need to tune the prefactors $\tilde{A}$ and $|W_0|$. Both examples feature $m_u\simeq 50$ TeV, $m_\v \sim \mathcal{O}(5\times 10^5)$ GeV and $m_{3/2} \sim \mathcal{O}(10^{10})$ GeV.

\begin{table}[h!]
\centering
\begin{tabular}{c|c||c|c|c|c|c|c|c|c|c}
$g_s$& $\tilde{A}$ & $N_s$ & $M$ & $|W_0|$ & $\tau_1$ & $\v$ & $\tau_s$ & $V_0\,10^{108}\,M_P^{-4}$ & $A_s$ & $\xi$ \\
\hline
\hline
0.1 & 1  & 3 & 362 & 1 & $1.24\times 10^4$ & $2.74\times 10^8$ & 9.32 & 1.7 & 1.2 & 1.8 \\
\hline
0.3 & 2  & 1 & 121 & 1 & $4.17\times 10^3$  & $3.29\times 10^8$ & 3.33 & 1.3 & 0.83 & 2 \\
\hline
\end{tabular}
\caption{Benchmark parameters that realize the EDE potential (\ref{VEDEC2LVS}) with $f=0.2\, M_P$ and $m_u \simeq 50$ TeV for K3 fibered CY models. We have used the LVS minimization relations (\ref{LVSmin}) with $\mathfrak{a}_s=2\pi/N_s$ and $\tilde{\mathfrak{a}}=2\pi/M$.}
\label{C2LVSExamplesNew}
\end{table}

Summarizing, our analysis shows that EDE can be realized in LVS with a $C_2$ axion whose potential is generated by gaugino condensates on D7-branes with non-vanishing world-volume fluxes. When the string coupling is small enough to trust the string loop expansion, matching the EDE scale requires $\mathcal{O}(100)$ D7-branes, if $\tilde{A}$ takes natural $\mathcal{O}(1)$ numbers. Swiss-cheese models with natural $\mathcal{O}(1)$ values of $|W_0|$ are characterized by $\v\sim\mathcal{O}(10^5)$ and $m_{3/2}\sim\mathcal{O}(10^{13})$ GeV, which can be lowered down at most to $m_{3/2}\sim \mathcal{O}(10^6)$ GeV by tuning $|W_0|$ (otherwise the volume modulus would cause cosmological problems). On the other hand, K3-fibered CY examples can realize EDE for larger values of the internal volume, $\v\sim \mathcal{O}(10^8)$, improving the control over the effective field theory. In turn, the resulting gravitino mass for $|W_0|\sim \mathcal{O}(1)$ is lower, $m_{3/2}\sim\mathcal{O}(10^{10})$ GeV.

\subsubsection{Gaugino condensation on D5-branes}
\label{sec:EDSLVSED1}

As explained in Sec. \ref{ED1D5}, $C_2$ axions can develop a potential also due to non-perturbative corrections to the K\"ahler potential arising from ED1-instantons or gaugino condensation on D5-branes. However, similarly to situation of the $C_4$ axions, this case implies a severe tuning on the prefactors of the non-perturbative effects to match the correct EDE scale. This fact is related to consistency of the model with the weak gravity conjecture as explained in \cite{Cicoli:2021gss}. We shall therefore be brief in the description of this case.

Focusing on the case where these non-perturbative effects correct the big modulus $\tau_b$, the K\"ahler potential and the superpotential would still be given by the standard LVS expressions (\ref{KLVS}) and (\ref{eqn:WLVS}) but now with the replacement:
\begin{equation}
\tau_b \to \tau_b - \gamma (G+\bar{G})^2+e^{-\tilde{\mathfrak{a}}t_b/ \sqrt{g_s}}\b{A_1\, \Re [e^{- \tilde{\mathfrak{a}}k\mathfrak{f} G}]+ A_2\, \Re [e^{- 2\tilde{\mathfrak{a}}k\mathfrak{f} G}]+ A_3\, \Re[ e^{-3\tilde{\mathfrak{a}}k\mathfrak{f} G}]} 
\label{VolED1}
\end{equation}
where $\tilde{\mathfrak{a}} = 2\pi/M$ and $G=\bar{S} b + \i c$, as defined in Sec.~\ref{Sec:Axions}. The scalar potential admits 3 contributions of the form
\begin{equation}
V = V_\text{LVS}(\v,\tau_s) + V_\text{EDE}(c) + V_b(\theta_b) \,,
\end{equation}
where $V_\text{LVS}$ and $V_b$ are given by \eqref{eqn:LVS_pot} and \eqref{eqn:Vb}, while the EDE potential is (for $b=0$)
\begin{equation}
V_\text{EDE} = V_0 \sb{\tilde{A}_1 \cos(\tilde{\mathfrak{a}}|k|\mathfrak{f}\, c)+\tilde{A}_2 \cos(2\tilde{\mathfrak{a}}|k|\mathfrak{f}\, c)+\tilde{A}_3 \cos(3\tilde{\mathfrak{a}}|k|\mathfrak{f}\, c)} \,,
\end{equation}
where we have set $A_i = \tilde{A} \tilde{A}_i$ ($i=1,2,3$) and
\begin{equation}
V_0 = \frac{3 \tilde{A}\,|W_0|^2 \tilde{\mathfrak{a}}^2}{2 g_s^2 \vol^2}\,e^{-\tilde{\mathfrak{a}} t_b/\sqrt{g_s}} \equiv \Lambda\,\tilde{A}\,e^{-\tilde{\mathfrak{a}} t_b/\sqrt{g_s}} \,.
\label{V0ED1}
\end{equation}
Note that the potential $V_b$ for the axion $\theta_b$ is decoupled from the EDE dynamics, and so $\theta_b$ can be safely set to zero, as in a standard LVS model. Using the canonically normalised field defined in \eqref{eqn:canon_c2}, we obtain in the potential the term
\begin{equation}
\cos\left(\tilde{\mathfrak{a}}|k|\mathfrak{f}\, c\right) = \cos\left(\tilde{\mathfrak{a}}|k|\mathfrak{f} \sqrt{\frac{\tau_b}{6\gamma}}\,\varphi\right) \equiv \cos\b{\frac{\varphi}{f}} \,,
\end{equation}
from which we can obtain the decay constant of the EDE field $\varphi$
\begin{equation}
f = \frac{1}{\tilde{\mathfrak{a}}\mathfrak{f}}\sqrt{\frac{3 g_s}{2 |k| \tau_b}}
= \frac{1}{\mathfrak{f}}\sqrt{\frac{3 g_s}{8\pi^2 |k|}}\frac{M}{\sqrt{\tau_b}}\,.
\end{equation}
Obtaining $f= 0.2$ in Planck units for $g_s\sim \mathcal{O}(0.1)$ and $\tau_b\gtrsim\mathcal{O}(100)$, clearly requires $M\gtrsim\mathcal{O}(30)$, suggesting that gaugino condensation on D5-branes is better than ED1-instantons which would anyway be volume-suppressed if both effects are present. Moreover, for $\mathfrak{f}=|k|=\tilde{k}=1$ and $f=0.2\,M_P$, the EDE scale becomes
\begin{equation}
V_0 = \Lambda\,\tilde{A}\,e^{-\frac{1}{\mathfrak{f}}\sqrt{\frac{3}{\tilde{k}|k|}}\left(\frac{M_P}{f}\right)} M_P^4 \sim 10^{-4}\,\Lambda\,\tilde{A}\, M_P^4\,,
\end{equation}
which shows that $V_0 \sim 10^{-108}\,M_P^4$ can be obtained only by tuning $\tilde{A}$ to exponentially small values, in complete analogy with the $C_4$ axion case (see (\ref{V0C4LVSnew})).

We therefore conclude that realizing EDE with $C_2$ axions and ED1/D5 non-perturbative effects requires always an exponential tuning of the prefactors. Given that we have shown instead that models with $C_2$ axions and ED3/D7 non-perturbative effects can realize EDE in a more natural way, it is important to check that ED3/D7 contributions to the scalar potential dominate over ED1/D5 effects. This is guaranteed if the non-perturbative corrections to $K$ are characterized by $\tilde{\mathfrak{a}}= 2\pi/M$, with $M\leq 2$, since in this case (\ref{V0ED1}) would give $V_0 \ll {\rm eV}^4$ for the values of $t_b$ and $g_s$ found in Sec. \ref{sec:EDELVSD7} which reproduce the correct EDE scale for ED3/D7 effects.

\section{Conclusions}

In this work we have performed a detailed analysis of the theoretical and phenomenological requirements to realize a viable EDE model \cite{Poulin:2018cxd} from string theory. We have focused on KKLT and LVS models in type IIB flux compactifications which are the best developed scenarios for moduli stabilization. Following the idea proposed in \cite{McDonough:2022pku}, we have tried to reproduce the EDE potential by exploiting 3 non-perturbative corrections to the effective action, considering both $C_4$ and $C_2$ axions. The outcome of our investigation is a set of working models, amongst which the most promising candidates to realize EDE in type IIB string theory are $C_2$ axions with a potential generated by gaugino condensation on D7-branes with non-zero world-volume fluxes. 
In this case the EDE scale and decay constant can be matched without tuning any of the underlying parameters and with the effective field theory approach under control. Let us explain in simple terms how we got to this conclusion by discussing the challenges outlined in Sec. \ref{Intro}:
\begin{enumerate}
\item {\bf Controlled de Sitter moduli stabilization:} As already pointed out, KKLT and LVS are well-studied frameworks for moduli stabilization. However, the main requirement, in both cases, to trust the low-energy supergravity approximation is that the internal volume $\v$ is stabilized at large values to keep control over $\alpha'$ corrections. More precisely, the dimensionful CY volume can be expressed as ${\rm Vol} = \v\,\ell_s^6$ (with $\ell_s=2\pi\sqrt{\alpha'}$), implying that the parameter controlling the $\alpha'$ expansion is $\epsilon_{\alpha'} =\alpha'\, {\rm Vol}^{-1/3}\simeq \v^{-1/3}$. In the simplest compactification with just a single K\"ahler modulus $\tau \simeq \v^{2/3}$, we need therefore to ensure that $\tau \gtrsim \mathcal{O}(100)$ so that $\epsilon_{\alpha'} \lesssim 0.1$. Writing the EDE decay constant $f$ in terms of the instanton action $S = 2\pi\tau/M$, with $M$ the number of branes, as $f\, S \simeq \lambda M_P$, as we did in Sec. \ref{Intro}, we easily see that matching $f\simeq 0.2\, M_P$ implies
\begin{equation}
\tau \simeq \frac{\lambda\,M}{2 \pi} \left(\frac{M_P}{f}\right) \simeq \lambda\,M\,.
\label{fRel}
\end{equation}
As found in \cite{Cicoli:2021gss}, $C_4$ axions with potential generated by ED3/D7 effects and $C_2$ axions with potential generated by fluxed ED1/D5 effects feature $\lambda\sim\mathcal{O}(1)$, in agreement with expectations from the weak gravity conjecture applied to axions \cite{Arkani-Hamed:2006emk, Rudelius:2015xta, Brown:2015iha, Hebecker:2015zss}. In this case, $\tau \gtrsim \mathcal{O}(100)$ can be achieved only by considering $M \gtrsim \mathcal{O}(100)$. On the other hand, $C_2$ axions with potential generated by fluxed ED3/D7 effects can lead to a violation of the weak gravity conjecture since they are characterized by $\lambda \simeq \sqrt{g_s \tau}$ \cite{Cicoli:2021gss}. In this case, (\ref{fRel}) reduces to $\tau \simeq g_s\,M^2$ which could give $\tau \gtrsim \mathcal{O}(100)$ for $M \gtrsim \mathcal{O}(30)$ if the string coupling is fixed (by an appropriate choice of background 3-form fluxes) at $g_s\lesssim \mathcal{O}(0.1)$ so that string perturbation theory does not break down. Hence, in all cases we are forced to consider situations with a relatively large number of branes. 
 
\item{\bf Decoupling of non-EDE modes:} This requirement is crucial to ensure that the EDE dynamics is not affected by any other field. The cleanest situation is therefore the one where all the non-EDE modes are stabilized at an energy scale which is higher than the EDE one. This observation implies that $C_0$ and $B_2$ axions are not well-suited to realize EDE since their shift symmetry is broken at perturbative level. Best candidates are instead $C_2$ and $C_4$ axions whose shift symmetry is broken only at non-perturbative level. More precisely, $C_2$ axion are in principle good EDE candidates in both KKLT and LVS models, while $C_4$ axions can play the role of the EDE field only in LVS models since in KKLT they would be as heavy as the corresponding saxions, thus inducing a cosmological moduli problem. 
 
\item {\bf Absence of fine-tuning:} Two levels of fine-tuning can be necessary to reproduce the EDE potential: a tuning to get the right periodicity, and an additional tuning to match the EDE scale. We found that $C_4$ axions in LVS require both tunings, and so appear to be the worst EDE candidates. $C_2$ axions with potential generated by ED1/D5 corrections to the K\"ahler potential can instead reproduce the required periodicity naturally but need tuning to obtain the correct EDE scale, and so do not seem to be optimal EDE fields. The best EDE candidates are instead $C_2$ axions with potential generated by fluxed ED3/D7 corrections to the superpotential since they can, in principle, avoid both tunings.

These results can be intuitively understood as follows. As explained in Sec. \ref{sec:EDELVSC4}, the EDE periodicity can be naturally realized only when the saxionic partner of the EDE axion is stabilized at zero. The saxion associated to $C_4$ controls the volume of a 4-cycle which cannot be set to zero since it would cause a deviation from the supergravity approximation. This implies that EDE models based on $C_4$ axion require tuning. On the other hand, the saxion associated to $C_2$ is the $B_2$ axion which is naturally fixed at $b=0$, implying that $C_2$ axions can realize the EDE periodicity in a more natural way. Regarding instead the matching of the EDE scale $V_0$ without any tuning of the UV parameters, as already explained in Sec. \ref{Intro}, this requires a violation of the weak gravity conjecture. In fact, (\ref{KeyRel}) with $f\simeq 0.2\,M_P$, becomes
\begin{equation}
V_0 \simeq A\,e^{- \lambda M_P/f}\,M_P^4 \simeq A\,e^{- 5 \lambda}\,M_P^4 \simeq 10^{-108}\,M_P^4\quad \text{for}\quad \lambda \simeq 50 \quad \text{if}\quad A\simeq 1\,.
\end{equation}
As we have already seen, $C_4$ axions with ED3/D7 effects and $C_2$ axions with fluxed ED1/D5 effects have $\lambda\sim\mathcal{O}(1)$, and so can match $V_0 \sim 10^{-108}\,M_P^4$ only by tuning $A$ to exponentially small values. On the contrary, $C_2$ axions with fluxed ED3/D7 effects feature
\begin{equation}
\lambda\simeq \sqrt{g_s \tau} \simeq \left(\frac{g_s\,M_P}{f}\right) \frac{3 M}{4\pi} \simeq 0.2\, M \quad \text{for}\quad g_s \simeq 0.2 \quad \text{and}\quad f\simeq 0.2\,M_P\,.
\label{KeyRel2}
\end{equation}
Hence, $\lambda\simeq 50$ can be achieved for $A\simeq 1$ and $M\sim\mathcal{O}(100)$, implying that $V_0$ can be realized without tuning only for gaugino condensation on D7-branes since ED3-instantons, if the corresponding action is written as $S=2\pi \tau/M$, can allow only for $M=1/p$ with $p\in \mathbb{N}$. In turn, such a relatively large number of D7-branes ensures that the effective field theory is fully under control since (\ref{fRel}) combined with (\ref{KeyRel2}) implies $\tau\simeq 0.2\,M^2 \sim \mathcal{O}(5\times 10^3)$. Such a large value of $\tau$ can be easily realized in LVS models, while it would imply a very low gravitino mass in KKLT scenarios where $m_{3/2}\sim e^{-2\pi\tau/N}\,M_P$ where $N$ is the number of D7-branes supporting the gaugino condensate that lifts the volume modulus. Requiring $m_{3/2}\gtrsim \mathcal{O}(1)$ TeV to ensure $m_\v \gtrsim 50$ TeV, implies $N\gtrsim\mathcal{O}(1000)$ which is very difficult to achieve in controlled CY orientifold compactifications with D7 tadpole cancellation. Thus, the only way-out in KKLT models to avoid such a huge number of D7-branes seems to involve again an exponential tuning of the prefactor $A$. 

This problem is absent in LVS models. As explained in Sec. \ref{sec:EDELVSD7}, $C_2$ axions with potential generated by 3 gaugino condensates on D7-branes with non-zero gauge fluxes can realize EDE without any tuning of the microscopic parameters. Two scenarios arise, depending on the topology of the underlying CY threefold. If the compactification space has a Swiss-cheese structure, $\tau$ is identified with the overall volume $\v\sim \tau^{3/2} \sim \mathcal{O}(10^5)$, leading to $m_{3/2}\sim\mathcal{O}(10^{13})$ GeV and $m_\v\sim \mathcal{O}(10^{10})$ GeV. If instead the internal space is a K3-fibered CY, the overall volume is controlled by 2 divisors $\v\simeq\sqrt{\tau_1}\tau_2$ and $\tau$ can be identified with the fiber modulus $\tau_1$. Thus, matching $f\simeq 0.2\,M_P$ fixes only $\tau_1\sim \mathcal{O}(5\times 10^3)$, but not $\v$, which can therefore be larger than in the Swiss-cheese case if moduli stabilization yields an anisotropic CY with $\tau_2\gg \tau_1\gg 1$. In fact, we have obtained $\v\sim \mathcal{O}(10^8)$, which improves the control over the effective field theory and leads to lower moduli masses: $m_{3/2}\sim\mathcal{O}(10^{10})$ GeV, $m_\v\sim \mathcal{O}(10^6)$ GeV and $m_u\sim 50$ TeV (where $u$ is the direction in the ($\tau_1$-$\tau_2$)-plane orthogonal to the volume mode $\v$).

\item {\bf Explicit Calabi-Yau realization:} Our analysis outlines the features that a globally consistent compactification should have to realize a viable EDE model. The next step, to build a full-fledged string model, would be to provide a rigorous description of the underlying Calabi-Yau threefold, orientifold involution and brane setup in a way compatible with tadpole cancellation. The setup should also explicitly realize 3 gaugino condensates on fluxed D7-branes wrapping different homologous representatives of the same divisor, or on a single stack of D7-branes where however some subsets of branes are differently magnetized. Such a detailed construction is beyond the scope of this paper and we leave it for future work.
\end{enumerate}

We conclude that our analysis establishes EDE as a viable model of string cosmology. While the model-building presented here is to some degree contrived, being designed to yield the $\left[1-\cos(\varphi/f)\right]^3$ EDE potential, it makes use of well-known and well-studied ingredients and does not rely on any exponential tuning of UV parameters for $C_2$ axions in LVS with gaugino condensation on fluxed D7-branes. In this sense, realizing our EDE models does not seem considerably harder than constructing other scenarios in string cosmology such as quintessence and inflation. 

This is an important step towards understanding and developing the predictions of the model: with a concrete model realization in hand, one may then investigate the interactions of the EDE field with other fields which generate complementary indirect signals of the EDE dynamics, e.g. gravitational waves from the coupling to gauge fields \cite{Weiner:2020sxn}, and may identify other ingredients in the model that play a cosmological role, such as an ultra-light axion component of dark matter. An additional interesting future direction of investigation is to unify these EDE constructions with other epochs of cosmic acceleration, namely with cosmic inflation or late-time dark energy. In the context of LVS, EDE seems particularly well suited to incorporating fiber inflation \cite{Cicoli:2008gp,Broy:2015zba,Cicoli:2016chb,Cicoli:2016xae, Cicoli:2017axo, Burgess:2016owb}, and the use of a $C_2$ axion to generate an observable level of chiral primordial gravitational waves \cite{McDonough:2018xzh}. There is also a natural compatibility of EDE presented here with fuzzy dark matter as presented in \cite{Cicoli:2021gss}, wherein one $C_2$ axion in the compactification could play the role of fuzzy dark matter and another the EDE. It would be interesting to revisit in this context some of the observables of fuzzy dark matter, such as dark matter substructure \cite{Alexander:2019qsh}. Moreover, an orthogonal approach would be to consider other possible EDE candidates, such as ultralight scalars that are composite states of fermions, following related work on ultralight dark matter, e.g. \cite{Maleknejad:2022gyf,Alexander:2020wpm,Alexander:2018fjp}. We leave these interesting avenues to future research.

\vspace{1cm}

\noindent{\bf Acknowledgments} \vspace{0.2cm}\\ 
\noindent We would like to thank Pramod Shukla and Roberto Valandro for useful conversations. We also thank Ralph Blumenhagen for relevant comments on a draft of this work. E.M. is supported in part by the National Sciences and Engineering Research Council of Canada via a Discovery Grant. This article is based upon work from the COST Action COSMIC WISPers CA21106, supported by COST (European Cooperation in Science and Technology).

\vspace{1cm}

\appendix

\section{LVS Moduli Stabilization with Anti-brane Uplift}
\label{AppLVS}

We will now concern ourselves with the potential \eqref{eqn:LVS_pot} after the stabilization of the axions. To find the minimum of the theory we first solve for $\tau_s$ by deriving $V$, finding
\begin{equation}
    e^{-\mathfrak{a}_s \tau_s} = \frac{3 |W_0| \sqrt{\tau_s}}{\mathfrak{a}_s A_s \v} \frac{\mathfrak{a}_s \tau_s-1}{4 \mathfrak{a}_s \tau_s-1} \simeq \frac{3 |W_0| \sqrt{\tau_s}}{4 \mathfrak{a}_s A_s \v}\, .
\end{equation}
Moreover, in order to simplify the equations, we perform the following change of variables
\begin{equation}
\psi \equiv \mathfrak{a}_s \tau_s \,,\qquad \v = \beta\,\sqrt{\frac{\psi}{\mathfrak{a}_s}}\,e^{\psi}\left(\frac{1-1/\psi}{1-1/(4\psi)}\right)\,,\qquad \beta =  \frac{3 |W_0|}{4 \mathfrak{a}_s A_s} \,,
\end{equation}
which implies $\psi\simeq \ln \v$. Thus, at leading order in $\psi \gg 1$, the scalar potential looks like
\begin{equation}
V(\psi) = -\frac{3\,|W_0|^2}{2\beta^3}\,e^{-3\psi}\left[1-\left(\frac{\mathfrak{a}_s}{\psi}\right)^{3/2} \frac{\hat\xi}{2}\right]\left(1+\frac{9}{4\psi}\right) + \frac{M^2}{\beta^{4/3}}\left(\frac{\mathfrak{a}_s}{\psi}\right)^{2/3} \,e^{-4\psi/3} \left(1+\frac{1}{\psi}\right),
\label{Vpsi}
\end{equation}
where we kept the first correction in the $\psi \gg 1$ expansion. Solving for the minimum and requiring it to be Minkowski (i.e. $\p V = V = 0$) we find the minima conditions to be
\begin{eqnarray}
\frac{\hat\xi}{2} &=& \left(\frac{\psi}{\mathfrak{a}_s}\right)^{3/2}  -\frac{9}{10\mathfrak{a}_s} \left(\frac{\psi}{\mathfrak{a}_s}\right)^{1/2} \simeq \left(\frac{\psi}{\mathfrak{a}_s}\right)^{3/2} \equiv \tau_s^{3/2} 
\label{LVSMinCond1} \\
M^2 & = & \frac{27}{20} \frac{|W_0|^2}{\mathfrak{a}_s^{2/3} \beta^{5/3}\psi^{1/3}}\,e^{-5\psi/3} \simeq \frac{27}{20} \frac{|W_0|^2}{\mathfrak{a}_s}\,\frac{\sqrt{\tau_s}}{\v^{5/3}}\,,
\end{eqnarray}
where again we included only the first correction for $\psi\gg 1$. Note that (\ref{LVSMinCond1}), when substituted in (\ref{Vpsi}), would give a leading order cancellation and an AdS vacuum for $M=0$. Lastly, let us mention that a dS minimum can be achieved by allowing small shifts of $M$.

 \bibliographystyle{JHEP}
\bibliography{D7EDE.bib}
 
\end{document}